\documentclass{elsart}

\usepackage{icarus}

\usepackage{natbib}
\usepackage{aas_macros}
\usepackage{array}

\usepackage{hyperref}
\usepackage{csquotes}
\hypersetup{colorlinks=true,linkcolor=blue,citecolor=blue,urlcolor=blue}

\usepackage{adjustbox}
\usepackage{rotating}

\bibpunct{(}{)}{;}{a}{,}{,}

\usepackage{graphicx}

\begin{document}

\begin{frontmatter}

\title{\small The CH$_4$ cycles on Pluto over seasonal and astronomical timescales}
\author[a]{T. Bertrand}, 
\author[b]{F. Forget},
\author[a]{O.M. Umurhan},
\author[a]{J.M. Moore},
\author[c]{L.A. Young},
\author[c]{S. Protopapa},
\author[d]{W.M. Grundy},
\author[e]{B. Schmitt},
\author[f]{R.D. Dhingra},
\author[g]{R.P. Binzel},
\author[g]{A.M. Earle},
\author[a]{D.P. Cruikshank},
\author[c]{S.A. Stern},
\author[h]{H.A. Weaver},
\author[a]{K. Ennico},
\author[c]{C.B. Olkin},
\author[]{the New Horizons Science Team},

\address[a]{National Aeronautics and Space Administration (NASA), Ames Research Center, Space Science Division, Moffett Field, CA 94035, United States }
\address[b]{Laboratoire de M\'et\'orologie Dynamique, IPSL, Sorbonne Universités, UPMC Univ Paris 06, CNRS, 4 place Jussieu, 75005 Paris, France.}
\address[c]{Southwest Research Institute, Boulder, CO 80302, United States}
\address[d]{Lowell Observatory, Flagstaff, AZ, United States}
\address[e]{Université Grenoble Alpes, CNRS, CNES, Institut de Plan\'etologie et Astrophysique de Grenoble, F-38000 Grenoble, France}
\address[f]{Department of Physics, University of Idaho 875 Perimeter Drive, Moscow, ID 83843, United States}
\address[g]{Department of Earth, Atmospheric, and Planetary Sciences, Massachusetts Institute of Technology, Cambridge, MA 02139, United States}
\address[h]{Johns Hopkins University Applied Physics Laboratory, Laurel, MD, 20723, United States}

\begin{center}
\scriptsize
Copyright \copyright\ 2005, 2006 Ross A. Beyer, David P. O'Brien, Paul
Withers, and Gwen Bart
\end{center}

\end{frontmatter}

\begin{flushleft}
\vspace{1cm}
Number of pages: \pageref{lastpage} \\
Number of tables: \ref{lasttable}\\
Number of figures: \ref{lastfig}\\
\end{flushleft}

\begin{pagetwo}{}
Tanguy Bertrand \\
National Aeronautics and Space Administration (NASA), Ames Research Center, Space Science Division, Moffett Field, CA 94035, United States \\

Email: tanguy.bertrand@nasa.gov\\

\end{pagetwo}

\begin{abstract}
Pluto's surface is covered in numerous CH$_4$ ice deposits, that vary in texture and brightness, as revealed by the New Horizons spacecraft as it flew by Pluto in July 2015. These observations suggest that CH$_4$ on Pluto has a complex history, involving reservoirs of different composition, thickness and stability controlled by volatile processes occurring on different timescales.
In order to interpret these observations, we use a Pluto volatile transport model able to simulate the cycles of N$_2$ and CH$_4$ ices over millions of years.
By assuming fixed solid mixing ratios, we explore how changes in surface albedos, emissivities and thermal inertias impact volatile transport.
This work is therefore a direct and natural continuation of the work by \citet{Bert:18}, which only explored the N$_2$ cycles. 
Results show that bright CH$_4$ deposits can create cold traps for N$_2$ ice outside Sputnik Planitia, leading to a strong coupling between the N$_2$ and CH$_4$ cycles. 
Depending on the assumed albedo for CH$_4$ ice, the model predicts CH$_4$ ice accumulation (1) at the same equatorial latitudes where the Bladed Terrain Deposits are observed, supporting the idea that these CH$_4$-rich deposits are massive and perennial, or (2) at mid-latitudes (25$^\circ$-70$^\circ$), forming a thick mantle which is consistent with New Horizons observations.
In our simulations, both CH$_4$ ice reservoirs are not in an equilibrium state and either one can dominate the other over long timescales, depending on the assumptions made for the CH$_4$ albedo. This suggests that long-term volatile transport exists between the observed reservoirs.
The model also reproduces the formation of N$_2$ deposits at mid-latitudes and in the equatorial depressions surrounding the Bladed Terrain Deposits, as observed by New Horizons. 
At the poles, only seasonal CH$_4$ and N$_2$ deposits are obtained in Pluto's current orbital configuration. Finally, we show that Pluto's atmosphere always contained, over the last astronomical cycles, enough gaseous CH$_4$ to absorb most of the incoming Lyman-$\alpha$ flux. 
\end{abstract}

\begin{keyword}
Pluto\sep CH$_4$\sep paleoclimate\sep Modeling\sep GCM\sep glacier\sep volatile transport \sep\\
\texttt{http://icarus.cornell.edu/information/keywords.html}
\end{keyword}


\section{Introduction}
\label{sec:intro}

\subsection{Pluto's ices as observed by New Horizons in 2015}

In July 2015, our vision of Pluto changed as the New Horizons spacecraft revealed a frozen world with unprecedented landscapes in the Solar System \citep{Ster:15}. 
The first analysis of spectral data from the Linear Etalon Imaging Spectral Array (LEISA) instrument on-board New Horizons showed that Pluto's water ice bedrock is covered by volatile ices such as N$_2$, CH$_4$ and CO, except in some parts of the equatorial regions, covered only by dark tholins \citep{Grun:16}.
Detailed spectroscopic analysis then revealed a more complex volatile ice distribution with different types of ice mixtures on Pluto's surface \citep{Schm:17,Prot:17}. 

\begin{figure}[!h]
\begin{center} 
	\includegraphics[width=15cm]{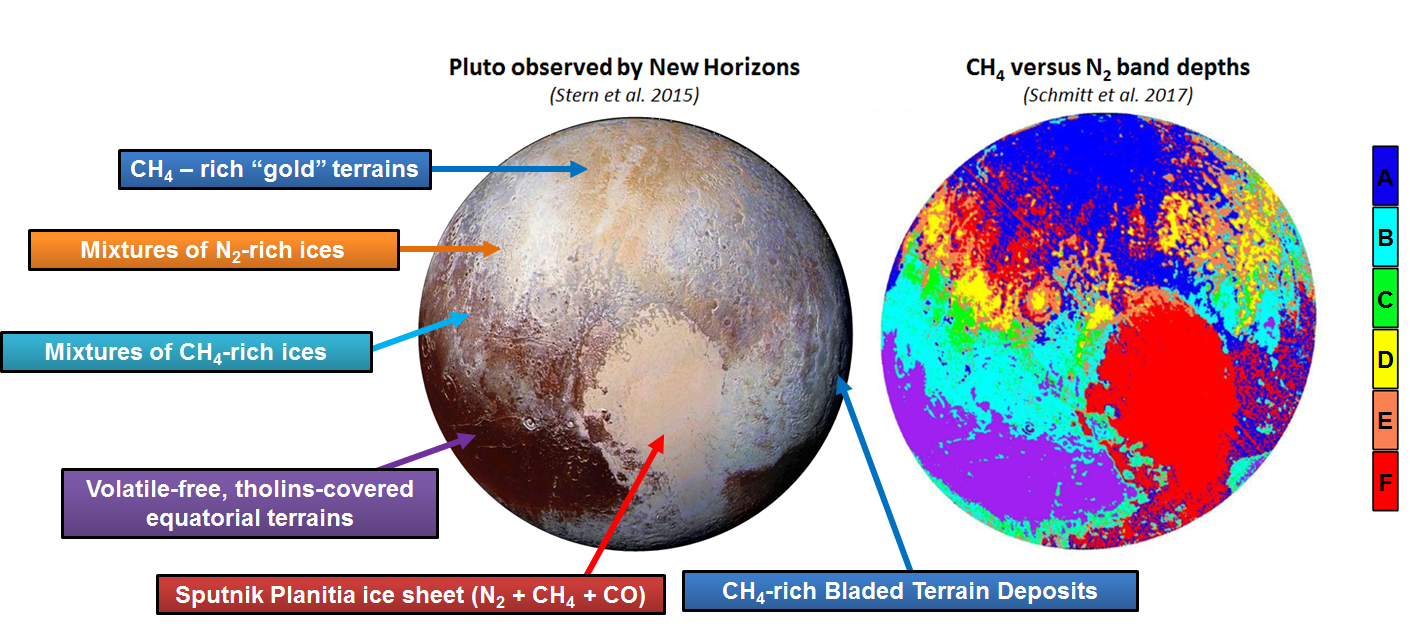}
\end{center} 
\caption{ The different types of terrains observed on Pluto (left) and the associated N$_2$-CH$_4$ mixtures (right). The color scale is described in details in \citep{Schm:17}: 
\textbf{A-B} relatively pure CH$_4$ ice, or N$_2$-rich ice with grains \textless~few cm and CH$_4$ \textgreater~1$\%$. \textbf{D-E} large N$_2$-rich grains (\textgreater~10~cm) with small amount of CH$_4$ (0.1-1$\%$). \textbf{C} N$_2$-rich ice with both medium N$_2$ grain size (\textless~few cm) and CH$_4$ \textless~1$\%$. \textbf{F} very large N$_2$ ice grains (\textgreater~20~cm) with CH$_4$ \textgreater~1$\%$. \newline
} 
\label{fig:distrib}
\end{figure}

The exact nature of the observed deposits is not easy to derive from these spectroscopic analyses, because they involve many parameters such as the abundance, dilution state, texture, or grain size, which are poorly constrained \citep{Schm:17}.
A first extraction of these parameters at the global scale is solved through sophisticated inversion of a Hapke's radiative transfer model of the LEISA data \citep{Prot:17}.
However, spectroscopic analyses of the surface do not provide information about the thickness of these deposits. The thickness can be inferred from geological insights using high resolution images from the LOng-Range Reconnaissance Imager (LORRI) instrument and albedo maps, that help distinguish a thin frost of ice from a massive deposit. A simplified global picture of the volatile ice reservoirs on Pluto is shown in \autoref{fig:distrib}.

The most prominent volatile ice deposit on Pluto's surface is a kilometer-thick ice sheet made of N$_2$ ice, mixed with CH$_4$ and CO, which is sequestered in Sputnik Planitia\footnote{The place names mentioned in this paper include a mix of officially approved names and informal names.}, the vast topographic basin at the anti-Charon longitude \citep{Ster:15,Grun:16}. This perennial glacier is the main reservoir of N$_2$ ice. 

Methane is detected almost everywhere in the northern hemisphere, with varying brightness, textures and mixtures (such as CH$_4$-rich ice and CH$_4$ diluted in N$_2$-rich ice), as highlighted in the available maps of the equivalent width of absorption in the Multi-spectral Visible Imaging Camera (MVIC) CH$_4$ filter \citep{Grun:16, Earl:18b} as well as in LEISA maps \citep{Prot:17,Schm:17}. 

The North Pole is covered by relatively pure and bright CH$_4$-rich ice, with an estimated Bond albedo higher than 0.7 \citep{Bura:17}. 



The mid-latitudes plains (25$^\circ$N-70$^\circ$N) are covered in mixtures of N$_2$-rich and CH$_4$-rich ices following a latitudinal trend \citep{Schm:17,Prot:17,Earl:18b}. The latitudes 55$^\circ$N-70$^\circ$N are enriched in CH$_4$, with few N$_2$-rich deposits mostly confined in the depressions. The latitudes 35$^\circ$N-55$^\circ$N are dominated by N$_2$-rich ices, while the latitudes 25$^\circ$N-35$^\circ$N are covered again mainly of CH$_4$-rich ices \citet{Prot:17}.
Interestingly, the CH$_4$-rich deposits in these regions seem to form a relatively thick mantle, maybe 100-1000 m, given the fact that they cover some craters and give to the surface a smooth aspect.

The equatorial regions (25$^\circ$S-25$^\circ$N) show a greater diversity of terrains in longitude, in terms of albedo \citep{Bura:17}, color \citep{Olki:17} and composition \citep{Schm:17}. Outside Sputnik Planitia, in the region of Tartarus Dorsa (East of Tombaugh Regio, 5$^\circ$S-28$^\circ$N), relatively pure CH$_4$-rich ice has been detected in the Bladed Terrain Deposits \citep[BTD,][]{Moor:18,Schm:17,Prot:17,Earl:18b}.
These terrains are characterized by parallel sets of steep ridges and sharp crests and are situated on high ground (above 2~km), which may indicate very massive CH$_4$-rich deposits, at least 300~m thick \citep{Moor:18}. 
They are relatively dark, with an estimated Bond albedo between 0.5-0.6 \citep{Bura:17}. 
Their distinctive texture on Pluto's maps suggests that they extend further east, from longitudes 210$^\circ$E to 40$^\circ$E, and further south, down to 25$^\circ$S \citep[see Fig. 3 in ][]{Olki:17,Moor:18}. They also seem to be interspersed with N$_2$-rich flat-floored bright plains located in the depressions and valleys of these regions.

Finally, the eastern part of Tombaugh Regio (between Sputnik Planitia and Tartarus Dorsa) is bright and also contains N$_2$-rich and CH$_4$-rich ices with N$_2$ mostly detected in the depressions \citep{Schm:17,Prot:17}.

These observations raise the following questions:
What drives the observed ice distribution and the diversity of N$_2$-rich and CH$_4$-rich terrains?
Are these reservoirs perennial (lasting for many Pluto years, e.g. glaciers) or seasonal (disappearing over one Pluto year, e.g. frosts)? 
How do they evolve over astronomical and seasonal timescales?
In this paper, we aim to provide answers to these questions by simulating the long-term evolution of N$_2$ and CH$_4$ ice, using the Pluto volatile transport model developed at the Laboratoire de M\'et\'eorologie Dynamique \citep[LMD,][]{BertForg:16, Bert:18}.

\subsection{Modeling of volatile transport on Pluto}

Global volatile transport models of Pluto have been used to explain how changes in insolation over the course of Pluto's orbit affect the surface and subsurface temperatures (which plays a key role in the Pluto environment), resulting in latitudinal variations of distribution of volatile ices \citep{Youn:93phd,HansPaig:96,Spen:97, Youn:12,Youn:13, HansPaig:15,Toig:15,Olki:15,BertForg:16,Bert:18}. 

In particular, \citet{BertForg:16} (hereinafter referred as BF2016) simulated the transport of N$_2$, CH$_4$ and CO ices over tens of thousands Earth years and obtained a seasonal cycle that reproduces, to first order, the ices distribution observed by New Horizons in 2015. They showed that N$_2$ ice inevitably accumulates inside Sputnik Planitia basin, because of an atmosphere-topography process: the N$_2$ surface pressure and condensing temperature are higher at the bottom of the basin than outside, and therefore the N$_2$ deposits in the basin are warmer and loose more energy by thermal infrared cooling, which is balanced by a stronger N$_2$ condensation at that location (so that enough latent heat is released to ensure that the surface remains at the equilibrium temperature).
In their simulations, the equatorial regions (apart from Sputnik Planitia) always remained warmer than at least one of the two poles (at any given time) and therefore no volatile condensed there, which explained the dark equatorial band of Pluto. 
They also showed that the lower volatility of CH$_4$ ice at Pluto's surface temperatures allows it to exist elsewhere than in the Sputnik Planitia ice sheet, forming frost even at locations where N$_2$ ice would immediately sublime. In their model, CH$_4$ ice seasonally covered both hemispheres, and, if its albedo was high enough, N$_2$ was able to condense on it and form a latitudinal band around 45$^\circ$N in 2015, in agreement with New Horizons observations (\autoref{fig:distrib}).
This impact of the ice albedo is also highlighted by \citet{Earl:18}, who suggest that runaway albedo variations are more efficient in the equatorial regions than at the poles, forming stark contrasts in albedo and volatile abundance.

However, the simulations performed in BF2016, and in most previous volatile transport modeling studies on Pluto, had two main limitations. First, they only considered small amounts of ice and were therefore not able to reproduce the formation of perennial glaciers.
For instance, the globally averaged volatile ice reservoir used in BF2016 is only of few millimeters, in order to reach a steady state over the annual timescale.
Secondly, they only focused on the 10$^4$ Earth years timescales, with the present orbital and obliquity parameters. Yet Pluto's high obliquity (currently around 119.6$^\circ$) varies by about 23$^\circ$ over a period of 2.8~millions of Earth years (2.8~Myrs herein refers to the astronomical timescale) while Pluto's longitude of perihelion regresses by 360$^\circ$ over 3.7~Myrs \citep{Dobr:97,Spen:97}. Both parameters impact the duration and intensity of the seasons, and the latitudes where volatile ices accumulate \citep{Binz:17,Earl:17}.

Recently, \citet{Bert:18} (hereinafter referred as B2018) improved the Pluto volatile transport model described in BF2016 by implementing the most recent topography data of Pluto, the variations of obliquity, longitude of perihelion and eccentricity with time, and by taking into account a realistic N$_2$ reservoir as well as a N$_2$ ice viscous flow model. Thanks to this modeling effort, they explored the cycles of N$_2$ over several astronomical cycles (up to 30~Myrs).
Their results explained many geological features of Sputnik Planitia, such as the evidence of recent and past glacial flow and erosion, the presence of sublimation pits in the southern portion of the ice sheet, as well as the brightness and composition of the surface ice. They also showed that large N$_2$ ice deposits can remain relatively stable and persist over tens of millions of years in the equatorial regions or over mid-latitudinal bands, in particular at low elevations.

\subsection{Objectives of this paper}

In this paper, we want to carry this work forward and explore the CH$_4$ cycles over astronomical and seasonal timescales with the latest version of the Pluto volatile transport model. In particular, our objectives are:
\begin{enumerate}
 \item To determine the latitudes where CH$_4$ ice tends to accumulate over millions of years.
 \item To investigate how the CH$_4$ albedo impacts the N$_2$ condensation-sublimation cycles and the latitudinal ices distribution.
 \item To compare our results with New Horizons observations of Pluto's surface \citep[e.g.][]{Ster:15,Grun:16,Moor:16}, explain the observed latitudinal distribution of volatile ices \citep{Schm:17,Prot:17}, infer the nature (perennial or seasonal) of the observed deposits, and discuss the possible scenarios for their formation.
\end{enumerate}
 
To fulfill these objectives, we adapt the Pluto volatile transport model described in B2018 so that it also takes into account the CH$_4$ cycle.
In Section~\ref{sec:model}, we detail the recent model improvements, the assumptions made and the simulation settings.
The results are presented in two independent sections. We first perform simulations with a global uniform CH$_4$ ice albedo (Section~\ref{sec:uniform}). Then, we perform simulations with two CH$_4$ albedo values (Section~\ref{sec:bladed}) depending on the latitude of the deposit, with a low value in the equatorial regions (old dark deposits) and high values in the mid-to-polar regions (bright deposits, above 30$^\circ$). Our results are summarized in Table~\ref{tab:results}, and discussed in Section~\ref{sec:discuss}.

\section{The Pluto volatile transport model}
\label{sec:model}

We use the latest version of the LMD Pluto volatile transport model, as described in BF2016, B2018 and \citet{Forg:17}.
The model settings are similar to those in B2018 except for the following changes, summarized in Table~\ref{tab:param}.

\subsection{General settings of the simulations}

The simulations of this paper are performed on a horizontal grid of 32$\times$24 points, which corresponds to a grid-point spacing of 7.5$^\circ$ in latitude by 11.25$^\circ$ in longitude (about 150~km at the equator).
We perform simulations over 30~Myrs using the paleoclimate mode, the N$_2$ ice viscous flow scheme and the ice equilibration algorithm described in details in B2018: the model is run over 5 Pluto years, then the annual mean ice rate of the last Pluto year is used to estimate the new amounts of ice over a paleo-timestep $\Delta$t and finally the topography is updated according to the new amounts of ice and the orbital parameters and the obliquity of Pluto are changed according to the new epoch t+$\Delta$t. The maximal change of obliquity within $\Delta$t must remains lower than the latitudinal resolution used. Here we use $\Delta$t~=~100,000 Earth years, that is about 400 Pluto orbits, allowing fast computing times and reasonable time resolution, with a maximal change of obliquity within that timeframe of about 3$^\circ$ \citep{Binz:17}.

Most of the atmospheric effects are neglected in the model (clouds, radiative effect of the atmsphere...). As in BF2016, volatile transport occurs through the atmosphere via a parametrization of the atmospheric circulation, using a characteristic timescale $\tau_{CH4}$~=~10$^7$~s (about four terrestrial months) to globally mix gaseous CH$_4$.

\begin{table}[h]
\begin{center}
\begin{tiny} 
\begin{tabular}{p{4cm}p{4cm}}
\textbf{Paleo-timestep} & $\Delta$t= 100,000 Earth years \\
\textbf{Reservoirs} \newline (global average) & $R_{N_2}$=300~m $^a$ \newline  $R_{CH_4}$=4~m $^b$ \\
\textbf{Albedo} & $A_{N_2}$= 0.7 \newline $A_{CH_4}$= 0.5-0.8 \newline $A_{bedrock}$= 0.1 \\
\textbf{Emissivity} & $\varepsilon_{N_2}$= 0.8 \newline $\varepsilon_{CH_4}$= 0.8 \newline $\varepsilon_{bedrock}$= 1 \\
\textbf{Thermal inertia} & TI=400-1500~SI $^c$ \\
\textbf{Atmospheric mixing timescale} & $\tau_{N_2}$= 1 s \newline $\tau_{CH_4}$= 10$^7$ s \\
\textbf{CH$_4$ ice mixing ratio in N$_2$} & 0.5 $\%$ \\
\hline
\multicolumn{2}{l}{$^{a}$\textit{Fills Sputnik Planitia up to 2500~m below the mean surface level}} \\
\multicolumn{2}{l}{$^{b}$\textit{In Section~\ref{sec:bladed}, an infinite reservoir is used}} \\
\multicolumn{2}{l}{$^{c}$\textit{SI= J s$^{-0.5}$ m$^{-2}$ K$^{-1}$}} \\
\hline   
\end{tabular}
\end{tiny}
\caption{Settings and surface conditions assumed in our simulations.
In this work we assess the impact of subsurface thermal inertia and CH$_4$ ice albedo.\newline}
\label{tab:param}
\end{center}
\end{table}

\subsection{Assumptions on the state of N$_2$ and CH$_4$ ice in the model}

On Pluto, CH$_4$ and N$_2$ easily mix together and are not expected to exist in perfectly pure state \citep{Traf:15,TanKarg:18}. Analysis of the New Horizons LEISA spectral observations of Pluto's surface has been performed using sophisticated spectral models and reveal complex mixtures in different amounts, also involving CO and contamination by tholins \citep{Grun:16,Prot:17,Schm:17}. Most of the volatile ice covering Pluto's surface seems to be dominated by N$_2$-rich:CH$_4$ (e.g. Sputnik Planitia) or CH$_4$-rich:N$_2$ (e.g. the north pole). Observations also suggest mixtures of both N$_2$-rich + CH$_4$-rich at some locations, but the exact nature of these deposits is uncertain for now because they could  fall into three distinct possible categories : (1) intimate mixture or intermolecular mutual attraction, at the grain scale, (2) geographic mixtures (N$_2$-rich + CH$_4$-rich could be observed at the pixel scale but not be mixed, striclty speaking, as they could be spatially disconnected at smaller scales), (3) stratification  (a thin layer of CH$_4$-rich ice could form at the top of a N$_2$-rich deposit).

The scenario of intimate mixtures suggests a perfect thermodynamic equilibrium of two types of cristals at significant depth, strickly following the binary phase diagram. This is typical of instantaneous thermodynamic equilibrium but should not apply to Pluto, which we believe is a non-equilibrium dynamical environment with continuous exchange of materials. Instead, CH$_4$-rich and N$_2$-rich ices may co-exist because of dynamical processes such as sublimation, which would lead to stratification (N$_2$ sublimation leaving CH$_4$ behind and leading to a CH$_4$-rich ice layer on top of the N$_2$ -rich ice).

However, the mechanisms controlling the formation and evolution of such N$_2$-rich + CH$_4$-rich mixtures remain largely unknown. With regards to this, the model presented in this paper is rather simple. As in BF2016, it does not  compute any evolution of ice mixing ratio. In the simulations, the surface is either volatile-free, covered by pure CH$_4$ ice or by N$_2$-rich:CH$_4$ ice. 
Pure CH$_4$ ice is an approximation for unsaturated CH$_4$-rich ice. We make the approximation that such a CH$_4$-rich ice behaves almost like pure CH$_4$ ice, in terms of temperature and vapor pressure at saturation of CH$_4$. It can form after sublimation of N$_2$ ice (in which it was trapped before) or directly on a volatile-free surface (its lower volatility than N$_2$ allows it to condense where N$_2$ would instantly sublime). For instance, CH$_4$-rich ice has been observed on top of mountains in the region of Chtulhu, where N$_2$ could not have condensed at first (because of the significantly high altitude and low albedo of these terrains). 

When both CH$_4$ and N$_2$ ices are present on the surface, we assume that CH$_4$ is diluted in a solid solution N$_2$:CH$_4$ with 0.5$\%$ of CH$_4$, as retrieved from telescopic observations \citep{Merl:15} and overall from the New horizons LEISA spectroscopic data \citep[see spectra g and j in Table 3 in ][]{Prot:17}. This modeled N$_2$-rich:CH$_4$ ice sublimes by conserving the 0.5$\%$ of diluted CH$_4$. In the next sections of this paper, we refer to this phase as N$_2$ ice.

The impact of N$_2$-rich + CH$_4$-rich mixtures is out of the scope of this paper, and we neglect their effect in the model. By doing so, we assume that this state corresponds to a short transient phase.
In the future, we plan to investigate further stratification mechanisms and implement them in the volatile transport model. To do this, experimental studies of these processes are also strongly needed.

\subsection{Surface properties}
\label{sec:modelsurf}

As in B2018, we use a reference N$_2$ ice albedo and emissivity which remain fixed to 0.7 and 0.8 respectively. The surface N$_2$ pressure simulated in the model is constrained by these values. 
The albedo and emissivity of the bare ground (volatile-free surface) are set to 0.1 and 1 respectively, which corresponds to a terrain covered by dark tholins such as Cthulhu. 
CH$_4$ ice emissivity is fixed to 0.8 in all simulations, but we explore different values of CH$_4$ ice albedo, ranging from 0.3 to 0.8 depending on the simulation (the reference albedo is 0.5). Note that CO is also transported by the model, but plays no role, as we assume it always ``follows'' N$_2$.

The reference seasonal thermal inertia (TI) of the subsurface is uniformly set to 800~SI, as in B2018. However we also performed simulations using a uniform soil TI set to 400 or 1200~SI (Section~\ref{sec:uniformsensib}), and simulations in which each type of terrain (water ice bedrock, N$_2$ or CH$_4$ ice) has its own TI ranging from 400 to 1200~SI (see Table~\ref{tab:results} and Section~\ref{sec:bladedsensibTI}). 



We also tested some scenarios in which the N$_2$ ice surface emissivity depends on its temperature and crystalline phase (e.g. simulation $\#$TI888$\_$050$\_$065$\_$phase). We assumed a minimal emissivity of $\varepsilon_{\alpha}$=0.3 when the ice is in its $\alpha$-phase, with a surface temperature T$_s$ below the transition temperature of T$_{\alpha-\beta}$=35.6~K and a maximal emissivity of $\varepsilon_{\beta}$=0.8 in its $\beta$-phase, based on the results from \citet{Stan:99}. We use a simple hyperbolic tangent function for the transition:

\begin{equation}
\varepsilon_{N_2}=\frac{1}{2}\left [  1+tanh(~3~(T_{\alpha-\beta}-T_s)~)\right ] ~(\varepsilon_{\alpha}-\varepsilon_{\beta}) + \varepsilon_{\alpha}
\end{equation}

As predicted by \citet{Stan:99}, the N$_2$ ice surface temperature in these simulations remains at the transition temperature of 35.6~K during most of Pluto's northern fall and winter, because the emissivity change leads to a negative feedback of surface temperatures.
The exchange of latent heat between both phases also leads to a negative feedback and plays a role in locking the temperature to the transition, but here we neglect this effect for sake of simplicity. 
 Note that the $\alpha$-phase of N$_2$ ice has never been observed in the outer solar system. 
Here we only test the case of an extremely low emissivity for N$_2$ ice in $\alpha$-phase, which enables us to assess the maximum possible effect of this change on Pluto's climate. 


\subsection{Reservoirs}
\label{sec:res}

All simulations are run with N$_2$ ice initially placed in Sputnik Planitia in a $\sim$10~km deep elliptical basin, as described in B2018. N$_2$ ice fills this modeled basin up to 2.5~km below the mean surface level, which corresponds to a global N$_2$ reservoir of $\sim$300~m. N$_2$ is allowed to condense and sublime everywhere on Pluto's surface, depending on the computed local surface thermal balance.

The simulations performed with a uniform CH$_4$ ice albedo (Section~\ref{sec:uniform}) use a global CH$_4$ reservoir of 2000~kg\,m$^{-2}$, which corresponds to 4~m, if we assume a CH$_4$ ice density of 500~kg\,m$^{-3}$ \citep{Leyr:16}.
This amount is low compared to the total reservoir of CH$_4$ ice expected on Pluto.
Indeed, the BTD only could correspond to a reservoir of 22~m on global average if we assume that they cover an equatorial surface equivalent to 10$^\circ$ in latitudes and 180$^\circ$ in longitudes and that they are only 500~m thick. This is a lower limit for CH$_4$ ice, since the BTD may be thicker and since other km-thick reservoirs may exist in the mid-latitudinal regions. 
However, running simulations with such a CH$_4$ ice reservoir is challenging, because the CH$_4$ condensation-sublimation rates are very low. Typically, one meter of CH$_4$ ice evolves over one million years. Therefore the total amount of CH$_4$ ice in our simulations is a trade-off between having the largest possible reservoir and reaching a steady state for ice distribution within tens of Myrs. 

Alternatively, the simulations performed with a dual CH$_4$ ice albedo (Section~\ref{sec:bladed}) are initialized with an infinite source of CH$_4$ ice at the location of the observed BTD in the equatorial regions. 


\subsection{Other settings related to the CH$_4$ cycle}

The model does not allow CH$_4$ ice to flow and the topography does not change according to the accumulation or loss of CH$_4$ ice (it only changes according to the variation of N$_2$ ice thickness). These choices are driven by the fact that (1) the CH$_4$ reservoir involved in the model is relatively low (see above) and (2) the CH$_4$ ice may be too rigid to lead to a significant glacial flow activity \citep{ElusStev:90,Moor:17,Moor:18}, as suggested by the steep slopes of the BTD's ridges (20$^\circ$), which is also why these terrains are described as massive deposits instead of glaciers.

Finally, for the simulations performed with a uniform CH$_4$ ice albedo (Section~\ref{sec:uniform}), we prohibit CH$_4$ to condense in the Sputnik Planitia N$_2$ ice sheet (we do not allow any CH$_4$ in SP). By doing this, we prevent the entire CH$_4$ reservoir to be trapped in Sputnik Planitia after several Myrs and we always conserve the same mass of CH$_4$ ice outside the basin. On Pluto, mechanisms or other sources must exist to maintain a certain amount of CH$_4$ ice outside Sputnik Planitia (see Discussions in Section~\ref{sec:discuss}). For instance, saturation of N$_2$-rich ice with CH$_4$, or formation of a CH$_4$-rich layer on top of N$_2$-rich ice are processes that could limit further condensation of CH$_4$ in SP, and maintain significant amounts of CH$_4$ ice outside SP. 

\section{Simulations with a uniform CH$_4$ ice albedo}
\label{sec:uniform}

In this section we investigate where CH$_4$ ice tends to accumulate over astronomical timescales.

\subsection{Initial state of the simulation}

The simulations are performed using the New Horizons topography data, starting with all N$_2$ ice filling a deep modeled Sputnik Planitia basin up to 2.5~km below the mean surface level, and with a 4-meter thick layer of CH$_4$ ice covering the entire globe.  
The simulation is performed with a unique albedo for CH$_4$ ice, set to 0.5, and a uniform thermal inertia of 800~SI. We then let the volatile ices evolve over 30~Myrs.

\subsection{Results: formation of massive equatorial deposits and mid-to-polar frosts of CH$_4$}

\begin{figure}[!h]
\begin{center} 
	\includegraphics[width=15cm]{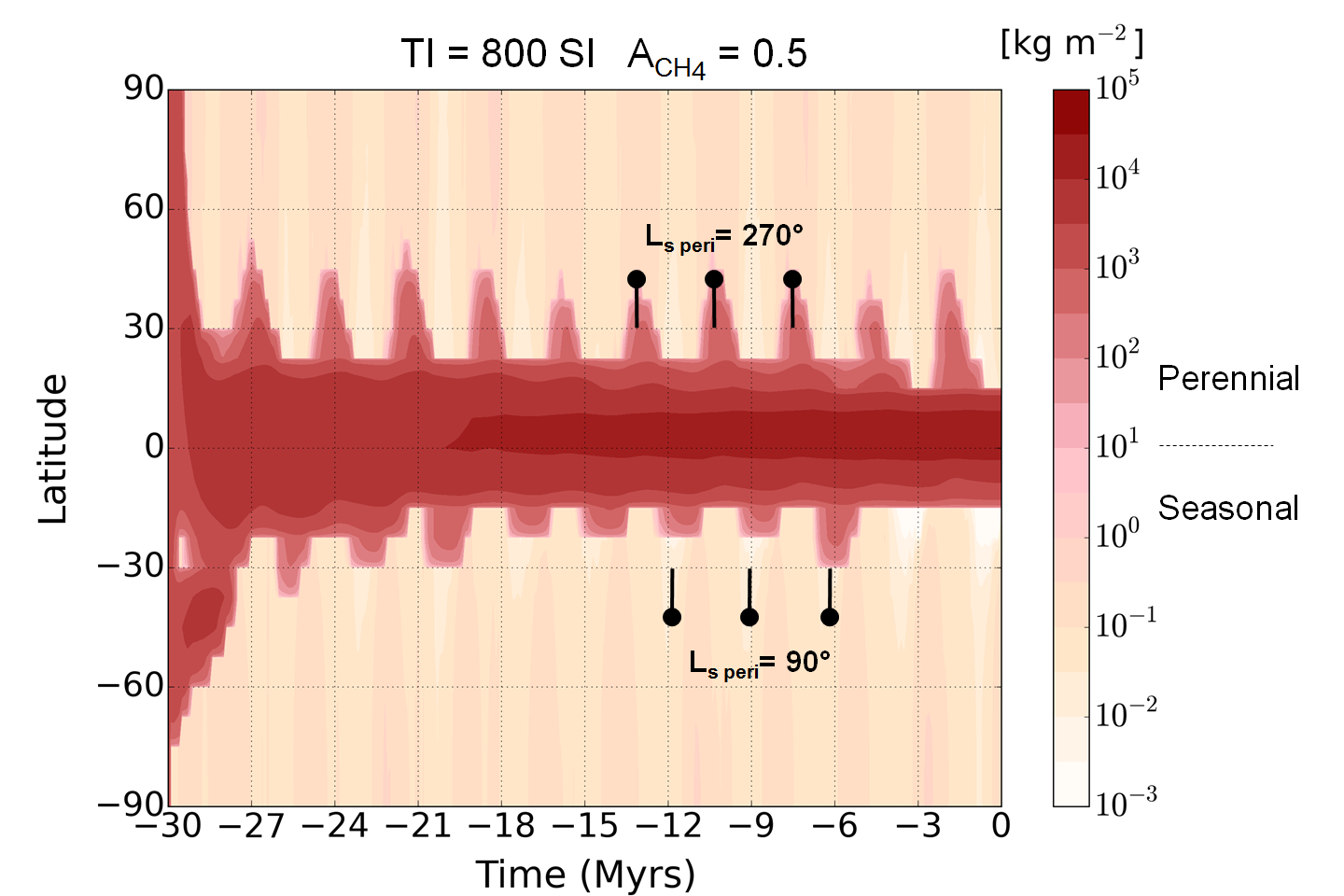}
\end{center} 
\caption{Evolution over 30~Myrs of the perennial and seasonal deposits of CH$_4$ ice on Pluto in our simulation with a unique CH$_4$ ice albedo of 0.5 (the values shown correspond to the zonal annual mean amount of CH$_4$ ice in kg\,m$^{-2}$, for the areas outside of Sputnik Planitia).  
CH$_4$ ice quickly accumulates in the equatorial regions, where it forms massive deposits. At higher latitudes, only thin seasonal CH$_4$ deposits form. The black vertical lines indicate the periods where L$_{s peri}$ is close to 90$^\circ$ and 270$^\circ$ (when seasonal asymmetries are the strongest). \newline
}
\label{fig:refuniform}
\end{figure}

\autoref{fig:refuniform} shows the evolution of the CH$_4$ ice distribution obtained over the last 30~Myrs. 
The ice quickly accumulates in the equatorial regions. Typically, after 10~Myrs, 20-40~m thick CH$_4$ deposits (1-2$\times$10$^4$~kg\,m$^{-2}$) are formed between 12$^\circ$S-22.5$^\circ$N (by accumulation of the CH$_4$ ice which was initially placed at the poles). 
The maximal net rates of sublimation are obtained at the poles. Above 30$^\circ$ latitude, 4 meters of ice can disappear within one astronomical cycle (2.8~Myrs). 
By extrapolation, if we had started with a global CH$_4$ ice reservoir of 100~m, we would have obtained $\sim$~1~km thick deposits in the equatorial regions after 50-100~Myrs. 

Why is CH$_4$ ice accumulating at the equator? 
Because of Pluto's high obliquity, ranging from 104$^\circ$ to 127$^\circ$ over 2.8~Myrs, the polar regions receive more solar flux than the equator on average. If one assumes medium to large soil thermal inertia, this leads to colder equatorial regions on average over several Myrs (see Fig. 4.B in B2018).
In addition to the obliquity cycle, Pluto's solar longitude of perihelion (L$_{s peri}$) oscillates over a period of 3.7~Myrs and leads to asymmetries in the seasons \citep{Dobr:97,Binz:17}. For instance, when Pluto's L$_{s peri}$ is close to 90$^\circ$, the northern polar latitudes undergoes a short and intense summer (close to the perihelion) and a long and intense winter (close to the aphelion) while the southern polar latitudes undergoes a long summer far from the sun and a short winter close to the sun, and vice versa when the orbital conditions are reversed (L$_{s peri}$ close to 270$^\circ$).
As a result, the northern hemisphere tends to be colder on annual average when Pluto's L$_{s peri}$ is close to 90$^\circ$, and warmer when Pluto's L$_{s peri}$ is close to 270$^\circ$ (see Fig. 4.A in B2018).

However, our results show that when L$_{s peri}$ is close to 90$^\circ$, the intense northern summer (occurring close to the perihelion) removes a significant amount of CH$_4$ ice at the north pole and during this epoch CH$_4$ ice accumulates at the equator and in the southern hemisphere. Conversely, when L$_{s peri}$ is close to 270$^\circ$, CH$_4$ accumulates at the equator and in the northern hemisphere. This is illustrated by \autoref{fig:refuniform} showing that CH$_4$ deposits extend to higher latitudes (20$^\circ$S and 45$^\circ$N) during the periods of asymmetric seasons (L$_{s peri}$ close to 90$^\circ$ or 270$^\circ$).  

On average over one astronomical cycle, the equatorial regions are a net accumulation zone of CH$_4$ ice, while the poles are a net sublimation zone. 
This result is consistent with New Horizons observations of the CH$_4$-rich BTD in the equatorial regions \citep{Moor:16,Moor:18} and supports the fact that they are thick and perennial CH$_4$ deposits.
 
\autoref{fig:refuniform} also shows that the modeled CH$_4$ deposits are not symmetric to the equator, as they tend to be more extended to the northern latitudes, in accordance with the observed latitudinal extent of the BTD (5$^\circ$S-25$^\circ$N). 
This is because during the last 70~Myrs, the L$_{s peri}$ value at high obliquity remained close to 90$^\circ$ and led to an asymmetry of insolation and surface temperatures which favors a slightly warmer south hemisphere (see details and Fig. 5 in B2018, and Discussions in Section~\ref{sec:discuss}).

Note that our modeling does not reproduce the “bladed” aspect of the BTD nor explain why they are mostly located in the eastern hemisphere, although it may be due to the fact that the dark tholin-covered surface in the western hemisphere (Cthulhu) prevents condensation of CH$_4$ (assuming that the BTD formed after Cthulhu). 
It is likely that this longitudinal asymmetry has dynamical origins and therefore it should be investigated by using 3D global climate models (which include a full 3D dynamical core).

Although thick deposits of CH$_4$ ice are not stable at the poles, thin CH$_4$ frost (\textless~1~mm) always form there seasonally, as illustrated by \autoref{fig:refuniform}. They form during fall-winter, when the equatorial deposits and the CH$_4$ frosts at the opposite pole (spring-summer) feed the atmosphere with gaseous CH$_4$. 
In BF2016, a similar result was obtained but because the simulations were not performed with a large enough reservoir of CH$_4$ ice, no thick deposit was obtained at the equator and the polar frosts disapeared in the early spring.


\subsection{Sensitivity to the reservoir, soil thermal inertia and CH$_4$ ice albedo}
\label{sec:uniformsensib}

Changing the initial spatial distribution of CH$_4$ ice (e.g. only at the poles, or over a specific longitudinal or latitudinal band) does not impact the results: CH$_4$ ice would still accumulate at the equator, with a slight extent to northern latitudes, while mm-thin deposits would form elsewhere during polar winter.
If we increase the CH$_4$ reservoir, then thicker CH$_4$ deposits are obtained at the equator.
The CH$_4$ mid-to-polar frosts remain qualitatively and quantitatively the same year after year because they are controlled by the location of the equatorial deposits.

\begin{figure}[!h]
\begin{center} 
	\includegraphics[width=15cm]{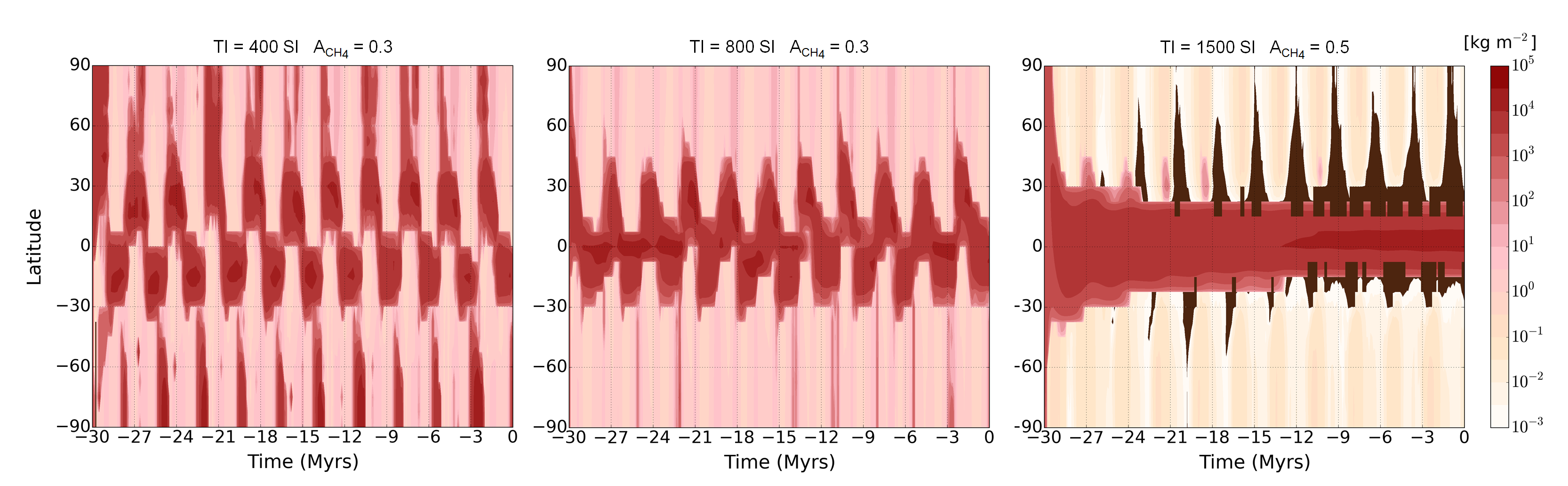}
\end{center} 
\caption{Same as \autoref{fig:refuniform} but assuming a thermal inertia TI=400~SI coupled with a CH$_4$ ice albedo A$_{CH_4}$=0.3 (left), TI=800~SI and A$_{CH_4}$=0.3 (middle), TI=1500~SI and A$_{CH_4}$=0.5 (right). 
The low TI case was performed with a lower albedo for CH$_4$ ice (0.3) in order to limit the effect of N$_2$ condensation on CH$_4$ ice (this effect is explored in Section~\ref{sec:bladed}). The brown color represents the dark, tholin-covered bedrock.\newline
} 
\label{fig:sensibuniform}
\end{figure}

Changing the albedo of CH$_4$ ice leads to major changes in the results. 
If we lower the uniform CH$_4$ ice albedo, the ice becomes warmer, resulting in higher sublimation rates and larger amounts of CH$_4$ to be transported to other sinks. In this case, we obtain more extended equatorial deposits and episodic thicker perennial deposits at the poles (see middle panel on \autoref{fig:sensibuniform}). 
If we increase the uniform CH$_4$ ice albedo, then CH$_4$ ice may become cold enough to trigger N$_2$ condensation on it, which strongly impacts both volatile cycles. This effect is explored in Section~\ref{sec:bladed}.

Results are also sensitive to the soil thermal inertia, as shown in \autoref{fig:sensibuniform}.
When a high thermal inertia is used (\textgreater~1500~SI), surface temperatures tend to be much warmer at the poles than at the equator on average over several Myrs (see B2018) and thus CH$_4$ ice accumulates closer to the equator, while less frost forms at the poles.
When a low thermal inertia is used (\textless~400~SI), the surface temperatures reach higher values in summer and lower values in winter and tend to be warmer at the equator than at the poles on average over several Myrs. 
In this case, large perennial reservoirs can form periodically in the mid-to-polar regions. 
However this case predicts, for present time, large reservoirs of CH$_4$ ice in the south hemisphere only, which is not consistent with the observations. 


\section{Simulations with darker CH$_4$ ice near the equator than at mid and high latitudes}
\label{sec:bladed}

CH$_4$ ice is known to play a complex role on Pluto's climate and volatile ices cycles since it can cold trap N$_2$ ice if its albedo is high enough \citep{BertForg:16,Earl:18}. In our model, the albedo of CH$_4$ ice is a key sensitivity parameter and we usually represent it by one value only, constant with time. However, on Pluto, the real value is not very well known and varies with time and space, because of different processes involving metamorphism effects, haze-particle settling/contamination (which serves as a darkening agent) and slight differences of composition \citep{Bura:17,Ster:15}. As an example, CH$_4$-rich ice is much brighter in the mid-to-polar than in the BTD \citep{Bura:17}. 

In this section, we intend to explore this sensitivity and better represent the cycle of CH$_4$ by considering two different albedos for CH$_4$ ice in the model, based on a criterion in latitude. In accordance with New Horizons observations, we assume that the mid-to-polar deposits are bright (albedo=0.65-0.8) and that the equatorial massive CH$_4$ ice deposits are dark (albedo=0.5-0.65). We also explore the impact of TI on the results (TI=400-1200~SI). The albedos of the volatile-free surface and of N$_2$ ice remain always fixed to 0.1 and 0.7 respectively. 

The results of this section are summarized in Section~\ref{sec:bladedsumarry} by \autoref{fig:resume} and Table~\ref{tab:results}. Note that all results shown for the current Pluto year are the outcome of 30~Myrs simulations.

\subsection{Initial state of the simulations}


N$_2$ ice is placed in Sputnik Planitia only. We place an unlimited CH$_4$ ice reservoir in the equatorial regions roughly at the locations of the observed BTD: 15$^\circ$S-15$^\circ$N, 140$^\circ$W-15$^\circ$E. The rest of the surface is initially volatile-free (assumed to be a tholin-covered water ice bedrock with an albedo of 0.1). We then run the simulations and let the volatile ices evolve over 30~Myrs.


\subsection{Reference simulation: accumulation of CH$_4$ ice at mid-latitudes}

Our reference simulation is named $\#$TI888$\_$050$\_$072, which means that is the simulation performed with a thermal inertia (TI) of 800~SI for N$_2$, CH$_4$ and water ices, an albedo for CH$_4$ equatorial deposits of 0.5 and an albedo for CH$_4$ mid-to-polar deposits of 0.72.

\subsubsection{The astronomical cycles of CH$_4$}

\begin{figure}[!h]
\begin{center} 
	\includegraphics[width=15cm]{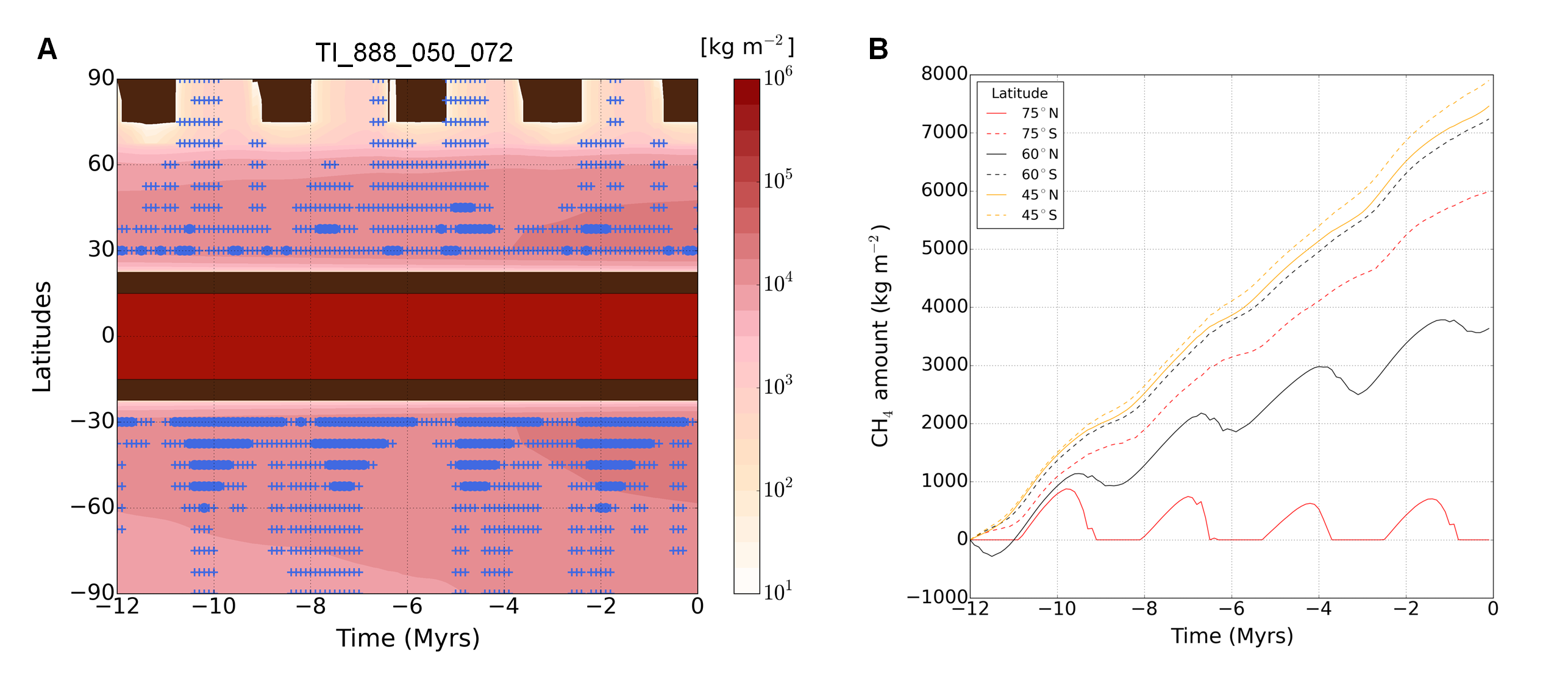}
\end{center} 
\caption{Simulation $\#$TI888$\_$050$\_$072. A. Evolution of the perennial CH$_4$ ice deposits: minimum amount of surface CH$_4$ ice (per Pluto year, over the last 12 millions of Earth years, that is $\sim$4 obliquity cycles). Seasonal deposits (frosts) are not shown. The dark-red band at the equator indicates the latitudes of the modeled BTD, which are an infinite source of CH$_4$ in the simulations. The dark-brown color represents the dark, tholin-covered bedrock.
The blue markers (cross and star) indicate the presence of N$_2$ ice perennial deposits, which are obtained at low elevations only (patchy deposits: cross) or over all longitudes (latitudinal band: star). 
B. Variation of the amount of surface CH$_4$ ice as shown in the panel A, for different latitudes and normalized at t=-12~Myrs. At the northern polar latitudes, perennial deposits form and disappear at each astronomical cycle. \newline
}
\label{fig:refbladedastro}
\end{figure}

\autoref{fig:refbladedastro} shows the evolution of the perennial CH$_4$ (in red) and N$_2$ (blue markers) ice deposits obtained over the last 12~Myrs. 
In this reference simulation, mid-to-polar CH$_4$ deposits form at higher latitudes than 30$^\circ$. Below, the equatorial regions remain volatile-free or covered by the modeled BTD, which is consistent with the observation of dark equatorial regions on Pluto (e.g Cthulhu).
These CH$_4$ deposits are cold enough to trigger N$_2$ condensation and allow the formation of perennial or seasonal N$_2$ deposits. Those located around 30$^\circ$N or 30$^\circ$S remain perennial at all times, while those located at higher latitudes remain perennial only during 1-2~Myrs during each astronomical cycle, as shown in \autoref{fig:refbladedastro}. Note that these perennial N$_2$ ice deposits remain less than 15~m thick, as shown in Table~\ref{tab:results}. 
At the North Pole, thin perennial CH$_4$ deposits (\textless~1~m) alternate with seasonal deposits over an astronomical cycle. Between 60$^\circ$N-75$^\circ$N, perennial CH$_4$ ice deposits (\textless~1~m) remain stable with time, while at northern mid-latitudes (25$^\circ$N-60$^\circ$N), a net accumulation of CH$_4$ ice is obtained (\autoref{fig:refbladedastro}.B).
The stability and accumulation of CH$_4$ ice at these latitudes is due to (1) the condensation of N$_2$ ice on top of the bright CH$_4$ ice, which is able to protect CH$_4$ from sublimation during most of the year and cold-trap even more CH$_4$, (2) the presence of infinite amounts of CH$_4$ ice at the equator (BTD), which continuously feed the atmosphere with gaseous CH$_4$. Indeed, the BTD are found to be a net sublimation zone of CH$_4$ ice over astronomical timescales, transporting CH$_4$ ice to the mid-to-polar regions. As shown in Table~\ref{tab:results} and \autoref{fig:refbladedastro}, $\sim$90~m of CH$_4$ ice are lost by the BTD and 30~m of CH$_4$ ice accumulated between 30$^\circ$N-60$^\circ$ after 30~Myrs. 

We performed the same simulations but starting at $t_0=$-100, -200 and -300~Myrs, in order to verify if the transport of CH$_4$ ice from the BTD to the mid-latitudes also occurs in different configuration for Pluto's orbit (the entire period of the cycle obliquity $+$ solar longitude of perihelion at maximum obliquity is 375 Myrs, as shown in Figure 5 in B2018). We obtained similar results.
Consequently, assuming that the Milankovitch cycles on Pluto remain stable with time, if we let this simulation evolve over 1~billion years, 3~km of CH$_4$ ice would have been transferred from the BTD to the mid-latitudes, where CH$_4$ ice would form a 1~km thick mantle. 
These results are consistent with the observations of New Horizons showing that the mid-latitudes are covered by a kilometer-thick mantle of volatile ice \citep{Howa:17}. They suggest that CH$_4$ has been accumulated there since hundreds of Myrs by the action of N$_2$ condensation-sublimation, and that the BTD have been losing significant amounts of CH$_4$ ice by sublimation. We further discuss this point in Section~\ref{sec:discuss}.

\subsubsection{ The current seasonal cycle of CH$_4$}

\begin{figure}[!h]
\begin{center} 
	\includegraphics[width=15cm]{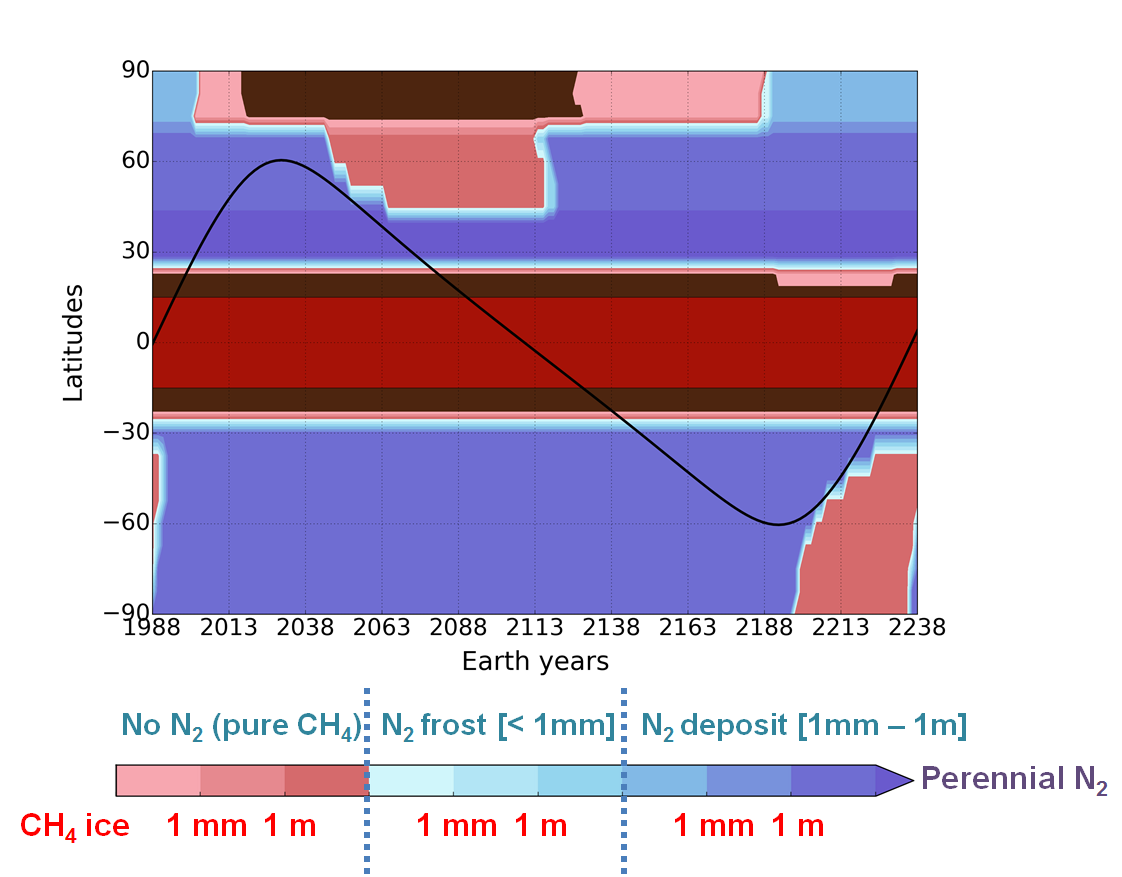}
\end{center} 
\caption{Reference simulation $\#$TI888$\_$050$\_$072. Evolution of the latitudinal distribution of the deposits over a current annual timescale (at longitude 0$^\circ$). The solid dark line shows the position of the subsolar point with time. Blue colors indicate the presence of N$_2$ ice and CH$_4$ ice while red colors indicate the presence of CH$_4$ ice only. The lighter the color, the thinner is the deposit. 
The dark-red band at the equator indicates the latitudes of the modeled BTD, which are an infinite source of CH$_4$ in the simulations. The dark-brown color represents the dark, tholin-covered bedrock.\newline
} 
\label{fig:refbladedseason}
\end{figure}

\autoref{fig:refbladedastro} shows that, at astronomical scale, present-time Pluto is in a period of its Milankovitch cycle where N$_2$ and CH$_4$ perennial deposits are not favored at high latitudes, compared to other periods (such as 2~Myrs ago for instance).
\autoref{fig:refbladedseason} shows the latitudinal distribution of N$_2$ (in blue, mixed with CH$_4$) and CH$_4$ (in red, no N$_2$) ice obtained over  a current Pluto year and after 30~Myrs of simulation. 

In the northern hemisphere, we can distinguish three different regions in latitudes. 
\begin{enumerate}
 \item  Around the North Pole (above 75$^\circ$N), CH$_4$ condenses as a thin frost (\textless~1~mm) during fall and triggers the condensation of N$_2$ ice on it during winter. When spring begins (Earth year 1988), the thin N$_2$ deposit (\textless~1~cm) starts to sublime. It disappears around year 2000, revealing a bright thin CH$_4$ frost that will last until 2017 before disappearing and leaving the dark substrate volatile-free during the entire summer. This is consistent with \autoref{fig:refbladedastro}.A, which predicts that there is no perennial deposit at the North Pole for the current epoch. 
 \item  At mid-latitudes (45$^\circ$N-75$^\circ$N), seasonal 1-m thick N$_2$ ice deposits cover CH$_4$ ice during most of the year, except during late spring and summer. During this period, N$_2$ ice sublimes from the pole and reveals the CH$_4$ ice mantle which starts to sublime as well. However, over the entire Pluto year, there is a net accumulation of CH$_4$ ice at these latitudes, leading to the formation of a thick mantle of CH$_4$ ice over Myrs, as shown in \autoref{fig:refbladedastro}.A. 
 \item  Between 30$^\circ$N-45$^\circ$N, a perennial latitudinal band of N$_2$ ice is obtained. 
\end{enumerate}

The southern hemisphere is covered by up to 1~m thick N$_2$ ice deposits during most of the year except during late summer, where it sublimes from the pole and reveals the CH$_4$ ice mantle. Around 1988, when southern fall begins, N$_2$ ice condenses and first covers the South Pole and then the mid-latitudes. A net accumulation of CH$_4$ ice occurs in the southern hemisphere over seasonal and astronomical timescales. 

\subsection{Sensitivity to the albedo of the equatorial CH$_4$ deposits}
\label{sec:bladedsensibalbeq}

\begin{figure}[!h]
\begin{center} 
	\includegraphics[width=15cm]{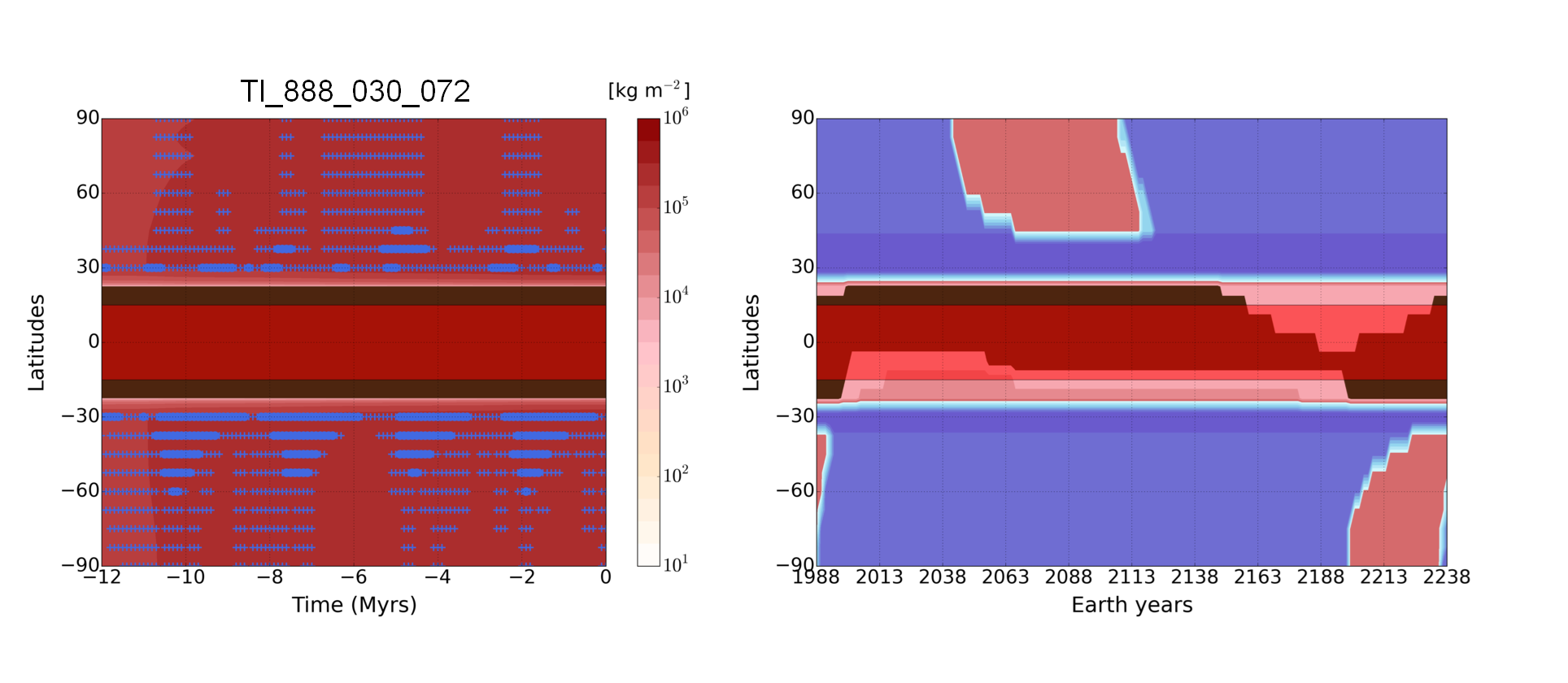}
\end{center} 
\caption{Simulation $\#$TI888$\_$030$\_$072. Same as \autoref{fig:refbladedastro} and \autoref{fig:refbladedseason} (legends are the same), except for an equatorial CH$_4$ ice albedo of 0.3 (very dark BTD). 
} 
\label{fig:bladed03}
\end{figure}

\begin{figure}[!h]
\begin{center} 
	\includegraphics[width=15cm]{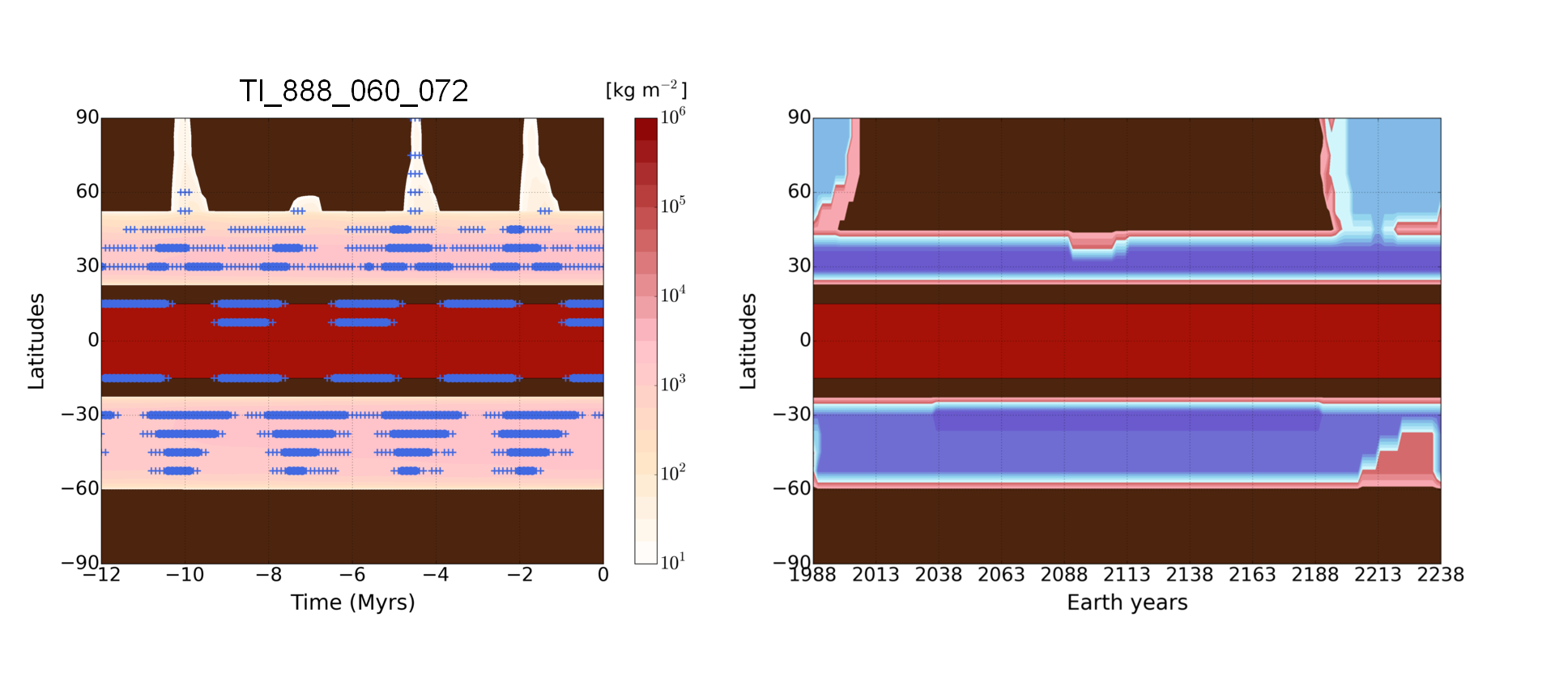}
\end{center} 
\caption{Simulation $\#$TI888$\_$060$\_$072. Same as \autoref{fig:refbladedastro} and \autoref{fig:refbladedseason}, except for an equatorial CH$_4$ ice albedo of 0.60 (relatively bright BTD).\newline 
} 
\label{fig:bladed065}
\end{figure}

Three different scenarios are obtained in our simulations if we increase or decrease the albedo of the modeled BTD (A$_{CH_4 eq}$):
\begin{enumerate}
 \item If A$_{CH_4 eq}$~$\ll$~0.6, the BTD are dark and warm enough so that they never trigger N$_2$ condensation. Lower albedo values lead to enhanced sublimation rates of CH$_4$ above these deposits and larger amounts of CH$_4$ ice transported to the mid-to-polar regions.  For instance, \autoref{fig:bladed03} shows the results obtained from simulation $\#$TI888$\_$030$\_$072, performed with an equatorial CH$_4$ albedo of 0.3. In this simulation, up to 1~km of CH$_4$ ice has been lost from the BTD after 30~Myrs (Table~\ref{tab:results}), and transported to the mid-to-polar regions. In this case, CH$_4$ is able to remain around the north pole and form perennial deposits at all times.  

 \item If A$_{CH_4 eq}$~$\approx$~0.6, condensation of N$_2$ occurs on the CH$_4$ equatorial deposits but only at low elevations (simulations $\#$TI888$\_$060$\_$072, $\#$TI888$\_$060$\_$080, $\#$TI888$\_$065$\_$072, $\#$TI888$\_$065$\_$080). The case of $\#$TI888$\_$065$\_$072 is shown in \autoref{fig:bladed065}. Up to 200-300~m thick N$_2$ ice deposits form in the low-elevated equatorial regions where CH$_4$ ice is present (see blue-star markers on \autoref{fig:bladed065}.A and $Max_{eqN_2}$ in Table~\ref{tab:results}). The high altitude BTD remain N$_2$-free and feed the atmosphere with gaseous CH$_4$ as they sublime, allowing the formation of perennial and seasonal CH$_4$ deposits at the poles. These mid-to-polar deposits are thinner than in the reference case because part of the equatorial source of gaseous CH$_4$ is trapped by N$_2$ ice. 

 \item If A$_{CH_4 eq}$~$\gg$~0.6, then the BTD are bright enough to trigger N$_2$ condensation at most altitudes. They are covered and cold trapped by 200-300~m of N$_2$ ice (see Table~\ref{tab:results}), mostly in the low-elevated regions since this is enough N$_2$ to flow downhill. These N$_2$ deposits remain relatively stable at such equatorial latitudes, as demonstrated by B2018 as well. In this case, the CH$_4$ sublimation at the equator is limited and thus there is not enough gaseous CH$_4$ available to form mid-to-polar deposits. As a result, Pluto's surface outside the equatorial regions remains volatile-free, which is not realistic (no simulation result is shown for this case).
\end{enumerate}

To summarize, Table~\ref{tab:results} shows that the amount of CH$_4$ ice lost by the modeled BTD over 30~Myrs is about 1~m, 10~m, 100~m or 1000~m assuming a CH$_4$ ice albedo of 0.65, 0.6, 0.5, or 0.3 respectively. 

\subsection{Sensitivity to the albedo of the mid-to-polar CH$_4$ deposits}
\label{sec:bladedsensibalbpoles}

Our results are also very sensitive to the albedo of the mid-to-polar CH$_4$ ice deposits. When this albedo is higher than 0.6, N$_2$ ice tends to condense on the CH$_4$ ice during fall-winter and form seasonal (few mm thick) or perennial deposits (up to 20~m thick), as shown for the reference case by \autoref{fig:refbladedastro} and \autoref{fig:refbladedseason}. If the mid-to-polar CH$_4$ ice albedo is lower than 0.6, then only thin seasonal frosts of CH$_4$ are obtained at the poles during fall-winter, as they do not trigger N$_2$ condensation. This case corresponds to the results obtained in Section~\ref{sec:uniform} and in BF2016. 

\begin{figure}[!h]
\begin{center} 
	\includegraphics[width=15cm]{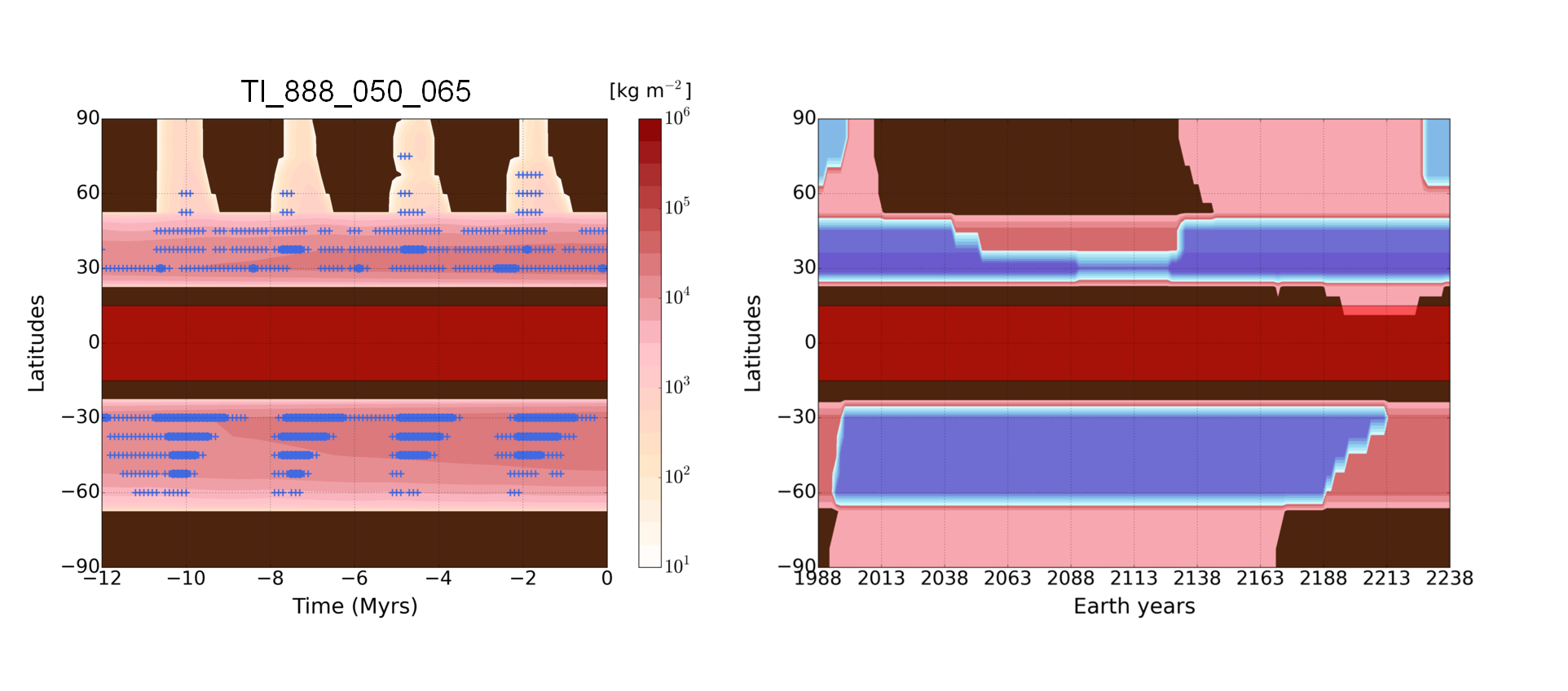}
\end{center} 
\caption{Simulation $\#$TI888$\_$050$\_$065. Same as \autoref{fig:refbladedastro} and \autoref{fig:refbladedseason}, except for a mid-to-polar CH$_4$ ice albedo of 0.65. 
} 
\label{fig:poles065}
\end{figure}

The higher the albedo of mid-to-polar CH$_4$ ice, the more N$_2$ condenses, at higher latitudes, and the longer it remains and traps CH$_4$ during Pluto's year. 
For instance, in our simulation using a mid-to-polar CH$_4$ ice albedo of 0.65 ($\#$TI888$\_$050$\_$065, \autoref{fig:poles065}), we obtain perennial N$_2$ deposits at 30$^\circ$N and very thin seasonal N$_2$ deposits at higher latitudes and in the southern hemisphere above 50$^\circ$S. As a result, CH$_4$ ice does not form thick deposits outside 50$^\circ$N-50$^\circ$S, but only seasonal frosts (e.g. the northern polar frost quickly disappears after 2013).

This is to be compared with the reference simulation (\autoref{fig:refbladedastro} and \autoref{fig:refbladedseason}), where the brighter mid-to-polar CH$_4$ ice albedo (0.72) triggers N$_2$ condensation at higher latitudes and leads to thick CH$_4$ ice deposits up to 70$^\circ$N and 90$^\circ$S, while the northern polar frost quickly disappears after 2017. 

\begin{figure}[!h]
\begin{center} 
	\includegraphics[width=15cm]{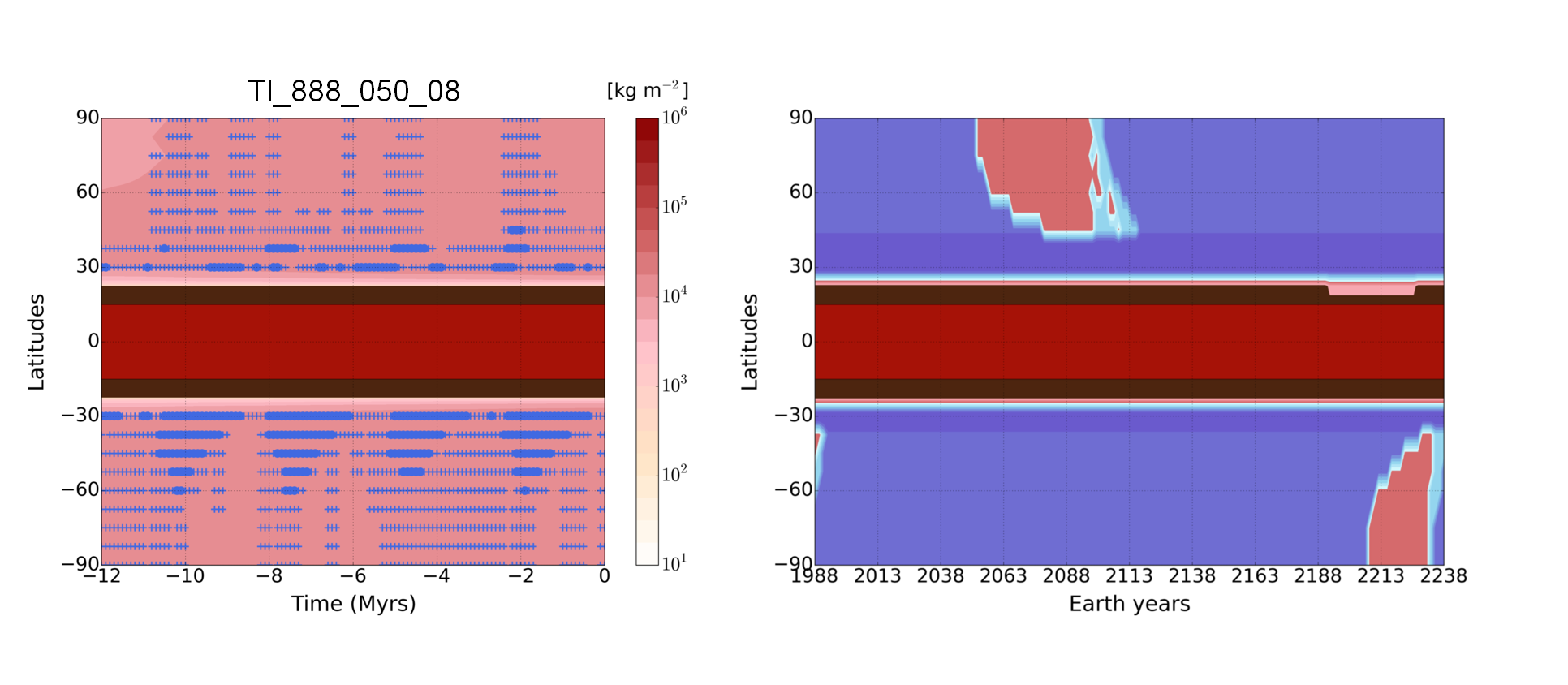}
\end{center} 
\caption{Simulation $\#$TI888$\_$050$\_$080. Same as \autoref{fig:refbladedastro} and \autoref{fig:refbladedseason}, except for a mid-to-polar CH$_4$ ice albedo of 0.8. \newline
} 
\label{fig:poles08}
\end{figure}

Finally, in the more extreme case of a mid-to-polar ice albedo of 0.8 ($\#$TI888$\_$050$\_$080, \autoref{fig:poles08}), perennial deposits of N$_2$ ice extend up to the pole during half of an obliquity cycle. During the current-year Pluto, N$_2$ ice covers the bright CH$_4$ ice up to the poles during most of the year, except during a short period in summer. In this simulation, long term accumulation of CH$_4$ ice is obtained everywhere outside the equatorial regions. 

\subsection{Sensitivity to N$_2$ ice phase emissivity}
\label{sec:bladedsensibphase}

The emissivity of N$_2$ ice in its $\alpha$-phase is less than that in its $\beta$-phase \citep{Stan:99,Lell:11b}.
 Here we tested the sensitivity of the results to the N$_2$ ice emissivity by assuming that it varies between $\varepsilon_{\alpha}$=0.3 in $\alpha$-phase and $\varepsilon_{\beta}$=0.8 in $\beta$-phase, as described in Section~\ref{sec:modelsurf}. 
This assumption has a strong impact on the N$_2$ cycle because the change of emissivity forces the ice surface temperature to remain at the transition temperature T$_{\alpha-\beta}$=35.6~K during most of Pluto's northern fall and winter. Consequently, higher annual mean N$_2$ ice surface temperature and surface pressure are obtained. 

\begin{figure}[!h]
\begin{center} 
	\includegraphics[width=15cm]{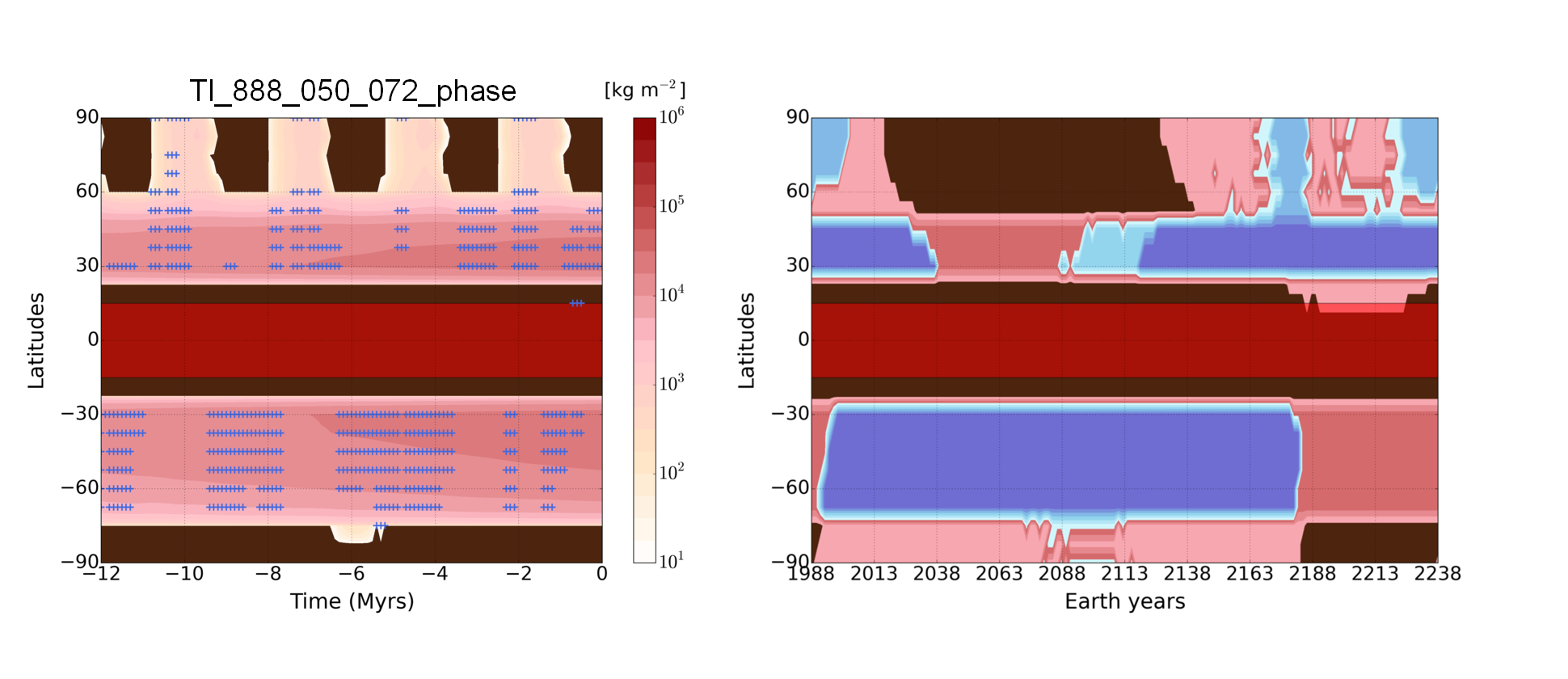}
\end{center} 
\caption{Simulation $\#$TI888$\_$050$\_$072$\_$phase. Same as \autoref{fig:refbladedastro} and \autoref{fig:refbladedseason}, except that the N$_2$ ice emissivity varies according to its phase ($\alpha$ or $\beta$) from $\varepsilon_{\alpha}$=0.3 to $\varepsilon_{\beta}$=0.8.\newline
} 
\label{fig:phase}
\end{figure}

\autoref{fig:phase} shows the results obtained for simulation $\#$TI888$\_$050$\_$072$\_$phase, which reproduces the reference simulation but taking into account the change of N$_2$ ice emissivity. The main differences with the results from the reference simulation (\autoref{fig:refbladedastro} and \autoref{fig:refbladedseason}) are: (1) N$_2$ ice does not form thick perennial deposits outside Sputnik Planitia. Over a current annual timescale, N$_2$ ice deposits in the mid-to-polar regions are only seasonal and disappear during summer. In the last 12~Myrs, perennial deposits are obtained only at low elevations, (2) No N$_2$ ice forms at the south pole during a current Pluto year, (3) CH$_4$ ice does not accumulate above 50$^\circ$N and below 70$^\circ$S. 

Consequently, the decrease of N$_2$ ice emissivity with temperature is an example of negative feedback which would limit the formation of N$_2$ deposits outside Sputnik Planitia. 
 
\subsection{Sensitivity to the thermal inertia of the different ices}
\label{sec:bladedsensibTI}

In the previous sections, simulations have been performed assuming a global uniform thermal inertia. In reality, the ices have different thermal inertia as it depends on the porosity of the material, the size of grains (larger grains lead to higher thermal inertia)... In this section, we allow each ice to have its own TI in the model, ranging from 400~SI to 1200~SI, and we explore the impact of these changes on the results (summarized in the second part of Table~\ref{tab:results}). 

For instance, simulation $\#$TI8412$\_$050$\_$065 has been performed with a thermal inertia of 800~SI for N$_2$ ice (“8”), 400~SI for CH$_4$ ice (“4”) and 1200~SI (“12”) for the water ice bedrock. In the model, TI evolves with time depending on the new thickness of the volatile ice on Pluto's surface. If a 1~m thick layer of CH$_4$ ice lies on water ice in the model, then the TI is set to 400~SI over the first meter of the subsurface and to 1200~SI below (in practice in the model the conductivity is modified to correspond to the required thermal inertia).

Changing the TI of N$_2$ ice (e.g. simulations $\#$TI488 or $\#$TI1288) does not significantly change the ice distribution. This is because to first order, the variation of the exchanged mass of N$_2$ between the surface and the atmosphere is independent of thermal inertia, as detailed in Section~2.2 in B2018. 

Changing the TI of CH$_4$ ice has also little impact on the ices distribution. Lower values of TI allows for colder CH$_4$ ice during winter, and eventually to slightly larger seasonal and perennial reservoirs of N$_2$ ice at the poles (simulations $\#$TI848,$\#$TI8412). The impact is also limited by the fact that the simulated CH$_4$ ice deposits are thin (few meter thick), because the initial reservoir is low (see Section~\ref{sec:res}). 

\begin{figure}[!h]
\begin{center} 
	\includegraphics[width=15cm]{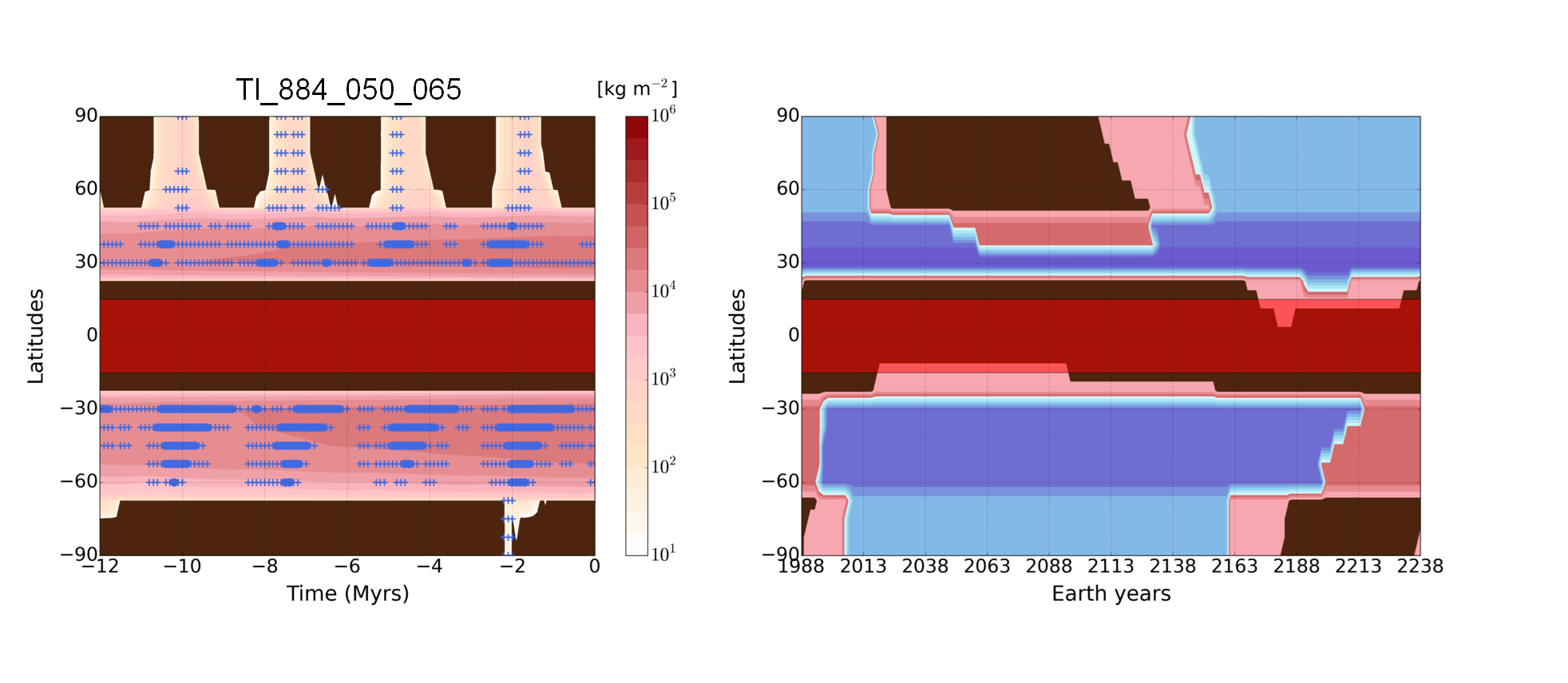}
\end{center} 
\caption{Simulation $\#$TI884$\_$050$\_$065. Same as \autoref{fig:refbladedastro} and \autoref{fig:refbladedseason}, except for a mid-to-polar CH$_4$ ice albedo of 0.65 and a TI for water ice of 400~SI. \newline
} 
\label{fig:poles065TIH2O400}
\end{figure}

Changing the TI of water ice has a significant impact on the thin volatile ice deposits, and therefore mostly at the poles. 
Low TI allows colder poles in winter and the formation of thicker CH$_4$ and N$_2$ deposits there (simulations $\#$TI884).
For instance, in the case of a mid-to-polar CH$_4$ ice albedo of 0.65 ($\#$TI884$\_$05$\_$065, \autoref{fig:poles065TIH2O400}), the northern polar N$_2$ deposit remains until years 2015-2020, and the CH$_4$ frost until 2025. This is to be compared with simulation $\#$TI888$\_$05$\_$065 (\autoref{fig:poles065}), where the northern frosts only last until 2000 and 2010 respectively, and where no N$_2$ condenses at the south pole. 
The effect is even stronger if we compare $\#$TI884$\_$05$\_$072 (\autoref{fig:poles072TIH2O400}) with the reference case TI888$\_$05$\_$072 (\autoref{fig:refbladedseason}). The N$_2$ polar deposit remains longer in northern spring and disappears after 2038, while the polar frost of CH$_4$ remains during the entire Pluto year.  

To summarize, our results are much less sensitive to TI (in the range 400-1200~SI) than albedos, although the TI of the water ice bedrock significantly impacts the distribution of the thin polar deposits. 

\begin{figure}[!h]
\begin{center} 
	\includegraphics[width=15cm]{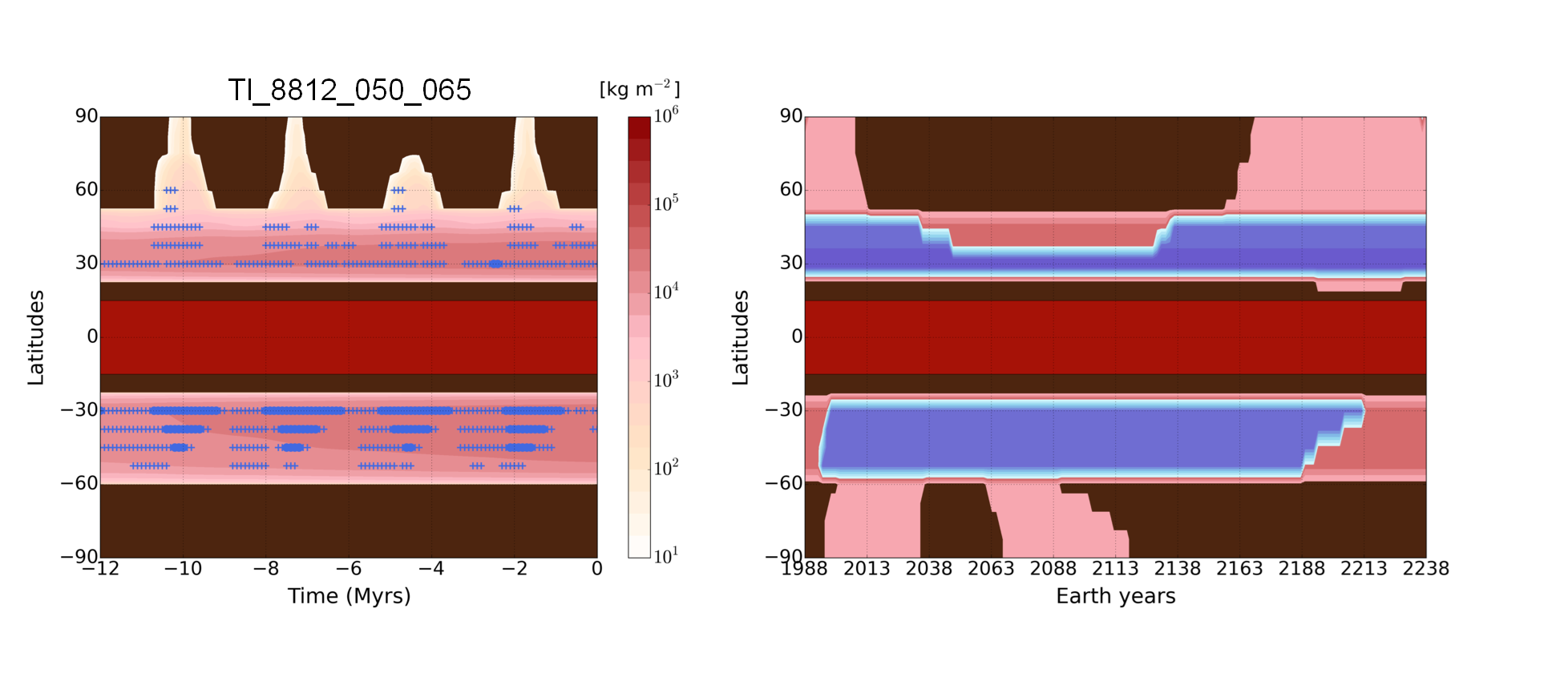}
\end{center} 
\caption{Simulation $\#$TI8812$\_$050$\_$065. Same as \autoref{fig:refbladedastro} and \autoref{fig:refbladedseason}, except for a mid-to-polar CH$_4$ ice albedo of 0.65 and a TI for water ice of 1200~SI. \newline
} 
\label{fig:poles065TIH2O1200}
\end{figure}

\begin{figure}[!h]
\begin{center} 
	\includegraphics[width=15cm]{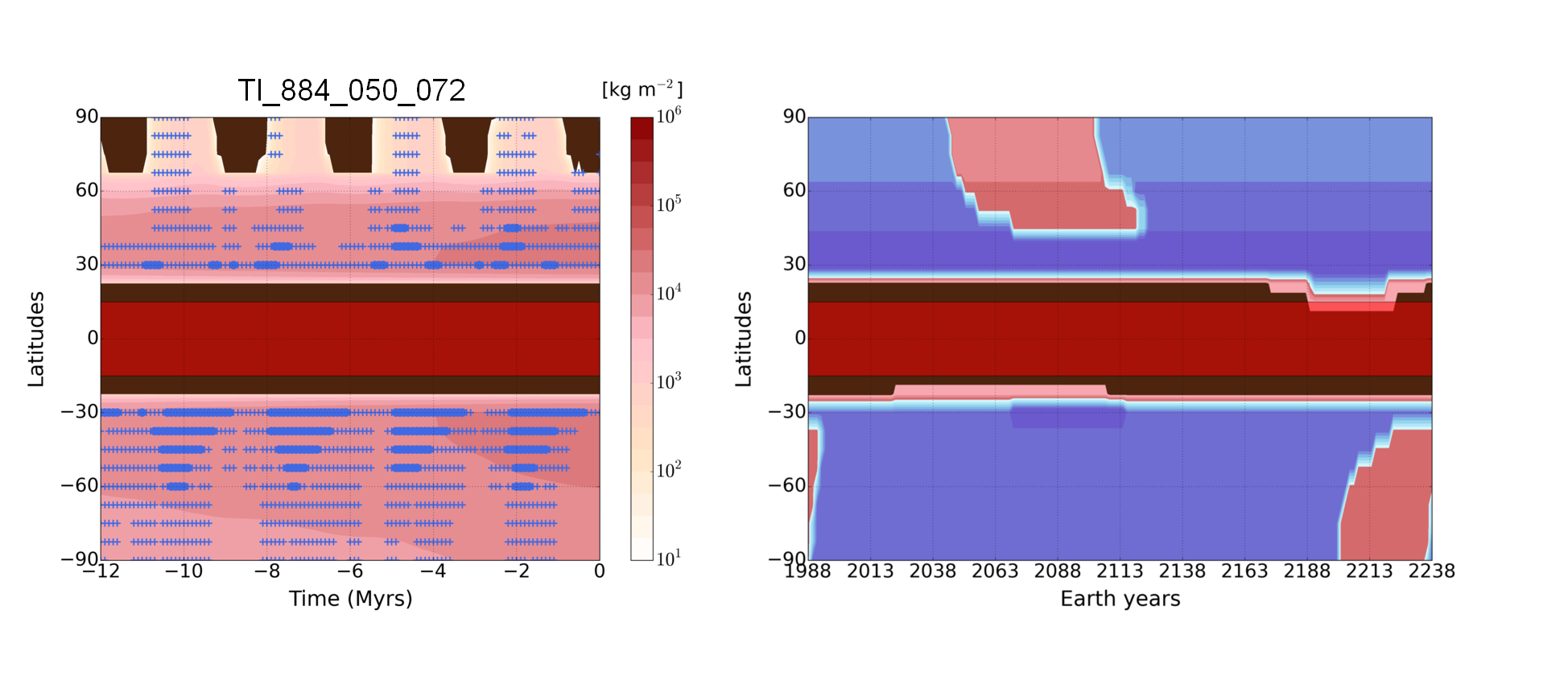}
\end{center} 
\caption{Simulation $\#$TI884$\_$050$\_$072. Same as \autoref{fig:refbladedastro} and \autoref{fig:refbladedseason}, except for a TI for water ice of 400~SI. \newline
} 
\label{fig:poles072TIH2O400}
\end{figure}

\begin{figure}[!h]
\begin{center} 
	\includegraphics[width=15cm]{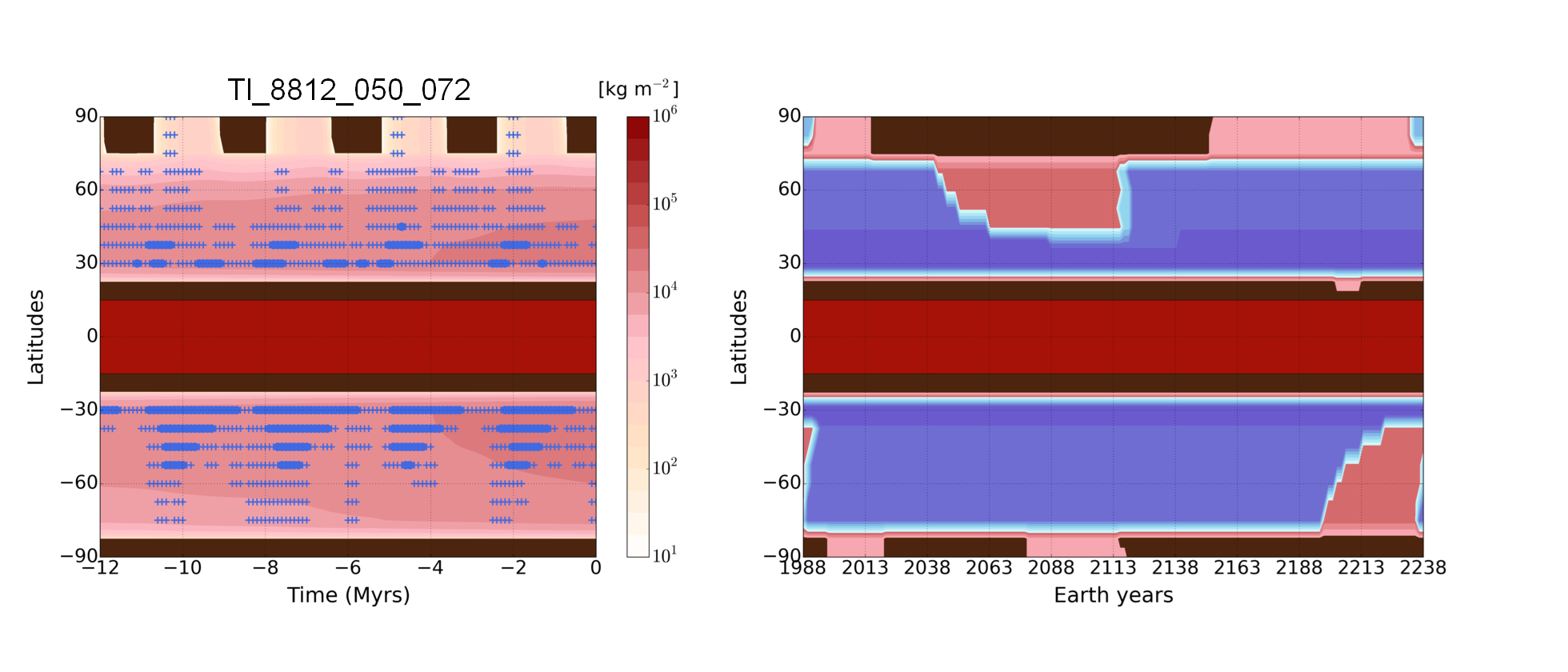}
\end{center} 
\caption{Simulation $\#$TI8812$\_$050$\_$072. Same as \autoref{fig:refbladedastro} and \autoref{fig:refbladedseason}, except for a TI for water ice of 1200~SI.  \newline
} 
\label{fig:poles072TIH2O1200}
\end{figure}

\subsection{Summary of simulation results}
\label{sec:bladedsumarry}

\begin{figure}[!h]
\begin{center} 
	\includegraphics[width=15cm]{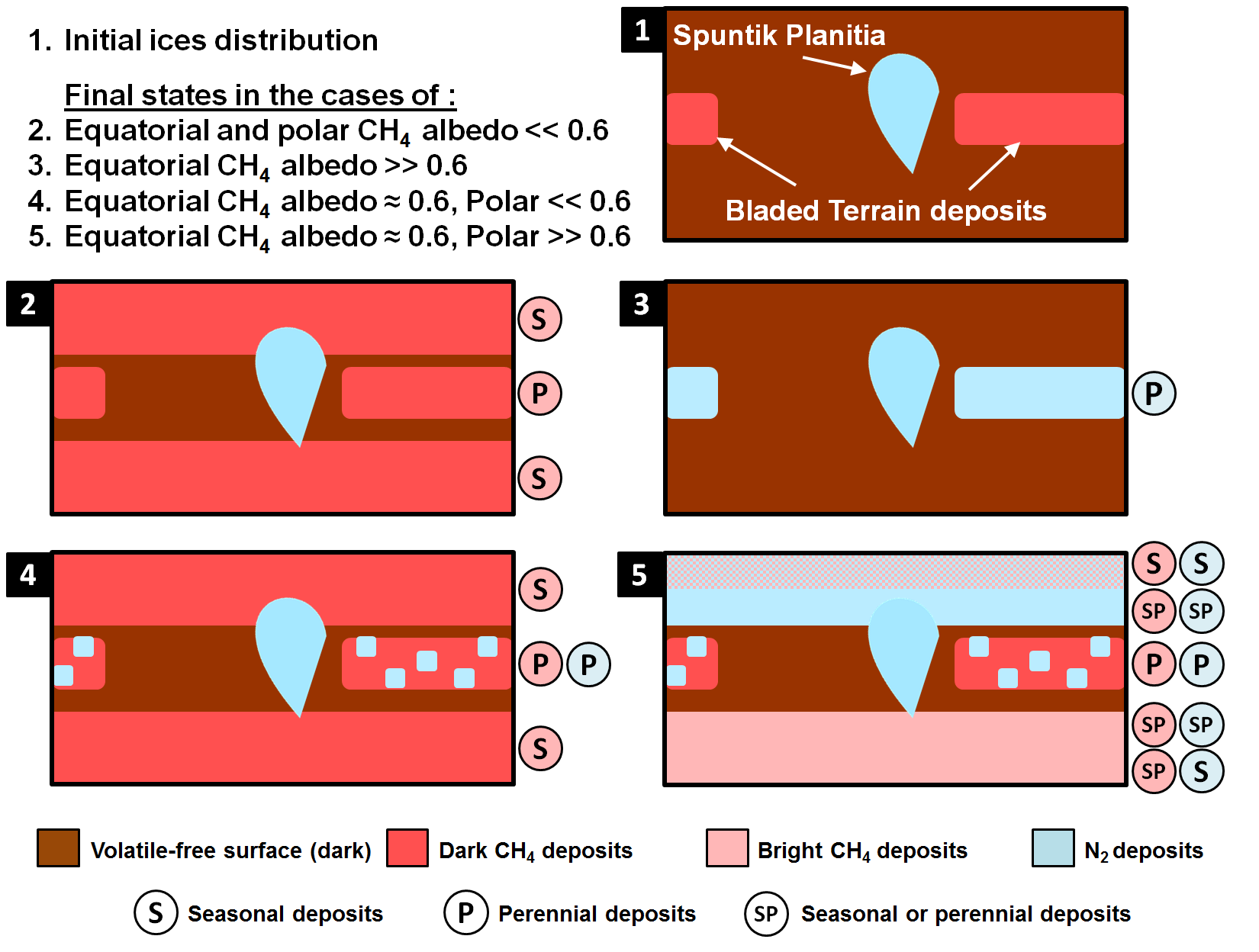}
\end{center} 
\caption{Summary of the simulation obtained in Section~\ref{sec:bladed}. \newline
} 
\label{fig:resume}
\end{figure}

\autoref{fig:resume} gives an overview of the different simulations performed in this Section~\ref{sec:bladed} and the different ices distributions obtained depending on the assumed albedo for CH$_4$ ice.
\autoref{fig:resume}.1 shows how the simulations were initialized: N$_2$ ice fills Sputnik Planitia while unlimited amounts of CH$_4$ ice are placed roughly at the location of the BTD. If the CH$_4$ ice albedo remains well below 0.6, only seasonal CH$_4$ frosts form at the poles, as shown in \autoref{fig:resume}.2. If the modeled BTD have an albedo well above 0.6 (\autoref{fig:resume}.3), then they become cold enough to trigger N$_2$ condensation. The N$_2$ ice deposits thus formed trap the CH$_4$ ice, which cannot feed the atmosphere with gaseous CH$_4$. As a result, there is no gaseous CH$_4$ left in the system and no frost can form at the poles, which remain volatile-free at all times. Using an albedo for the BTD around 0.6 allows the formation of N$_2$ ice deposits only in the depressions of these terrains, as shown in \autoref{fig:resume}.4. The high-altitude BTD remains N$_2$-free and feed the system with CH$_4$, allowing the formation of seasonal frosts at the poles. Finally, if the albedo of the mid-to-polar CH$_4$ deposits is set higher than 0.6 (\autoref{fig:resume}.5), then N$_2$ can condense and form thin seasonal deposits at the poles and larger deposits at mid-latitudes, which can be perennial or seasonal (up to few tens of meter thick, in particular in depressions). These N$_2$ deposits are able to trap large amounts of CH$_4$ ice, resulting in the formation of a thick mid-latitudinal mantle of CH$_4$ ice. 

\subsection{Surface pressure and CH$_4$ atmospheric mixing ratio}
\label{sec:pres}

\subsubsection{The peak surface pressure during northern spring}

Simulations from BF2016 predicted an evolution of surface pressure in accordance with the stellar occultation observations conducted from Earth since 1988. The threefold increase of pressure observed would result from N$_2$ ice heating when (1) Sputnik Planitia is most exposed to sunlight (shortly after the northern spring equinox in 1989) and (2) Pluto is close to the Sun. Their model also predicts that the atmospheric pressure should decrease in the following decades, after reaching its maximum around 2015, because of the orbitally-driven decline of insolation above Sputnik Planitia. 

Here, in this paper, although the general aspect of the annual evolution of surface pressure remains unchanged (see Figure~2.a in BF2016), the peak surface pressure occurs earlier than 2015 in many of our simulations. 
The main differences between the simulations of this paper and those from BF2016 are the presence of N$_2$ ice deposits outside Sputnik Planitia that slightly affect the evolution of pressure (by enhancing the global condensation or sublimation flow) and the better resolution of the Sputnik Planitia basin (BF2016 only assumed a circular crater). 

Our simulations show that the annual pressure peak occurs when the area of the sublimation source in the northern hemisphere becomes less than the area of the condensation sink in the southern hemisphere. For instance, in the reference simulation ($\#$TI888$\_$050$\_$072), the peak occurs in year 2000 when the northern polar deposit of N$_2$ disappears.

Simulations with a peak surface pressure occurring after 2010 are (1) the ones with a very high mid-to-polar CH$_4$ albedo, leading to both hemispheres covered by N$_2$ ice during northern spring (these cases are not realistic because they do not correspond to the ice distribution observed by New Horizons), (2) the ones with a thermal inertia for water ice around 400~SI ($\#$TI444, $\#$TI884, $\#$TI12124…). In these cases, the thin northern polar deposit of N$_2$ lasts until 2010-2015, (3) The ones without N$_2$ ice outside SP (like in BF2016). This scenario is only obtained when using in the model a “dark” mid-to-polar CH$_4$ albedo (less than 0.6). 

These results suggest that the southern hemisphere of Pluto is not entirely covered by N$_2$-rich ice, otherwise the peak surface pressure would have occurred much earlier than 2015 (a similar result is found in \citet{Youn:13,Olki:15}). 
At most, a thin mid-latitude band of N$_2$-rich ice (similar to that observed in the northern hemisphere) could be present in the southern hemisphere in 2015.

\subsubsection{Evolution of surface pressure over astronomical timescales}

In all simulations of this paper, the surface pressure obtained remains within few mPa-Pa, with a maximal value of 4~Pa over 30~Myrs. This is in the same range than the values obtained in B2018 (see Figure 16 of their paper, lines obtained with an albedo of 0.7 for N$_2$ ice). 
We could have expected higher values in the results of this paper because we obtained N$_2$ deposits at the poles. However, the increase of pressure due to their sublimation in summer is always balanced by the strong condensation flux at the opposite pole. 

\subsubsection{Evolution of CH$_4$ atmospheric mixing ratio over astronomical timescales}

\begin{figure}[!h]
\begin{center} 
	\includegraphics[width=15cm]{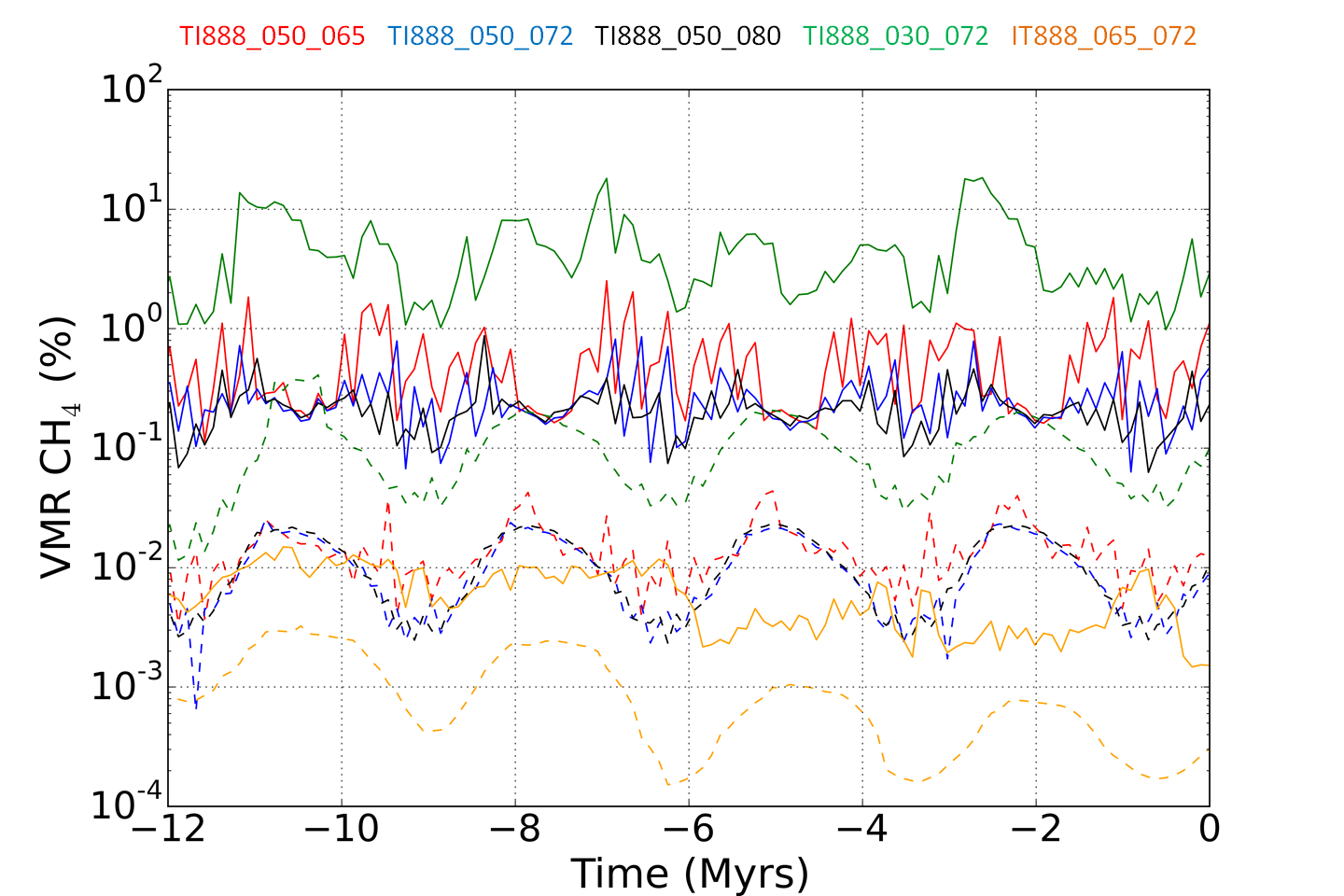}
\end{center} 
\caption{Evolution of the annual maximum (solid line) and minimum (dashed line) global mean atmospheric mixing ratio of CH$_4$ for different simulations of this paper.  \newline
} 
\label{fig:vmr}
\end{figure}

The atmospheric mixing ratio of CH$_4$ obtained is very sensitive to model parameters \citep{BertForg:16}, in particular those controlling the BTD (main source of gaseous CH$_4$). \autoref{fig:vmr} shows the annual maximum and minimum values obtained over astronomical timescales for different simulations. We note that (1) The CH$_4$ atmospheric mixing ratio remains within 0.001-1$\%$ in most of the simulations, including the reference case, (2) Higher values can be obtained for a lower equatorial CH$_4$ albedo (0.01-10$\%$ with an albedo of 0.3), (3) Lower values are obtained when N$_2$ condenses on the equatorial CH$_4$ deposits (10$^{-4}$-10$^{-2}\%$), (4) The lower the albedo of the mid-to-polar CH$_4$ deposits, the higher the concentration of CH$_4$ in the atmosphere (because of the higher equilibrum temperature and pressure of CH$_4$ and because there is less N$_2$ deposit forming and therefore more CH$_4$ ice available to feed the atmosphere with gaseous CH$_4$).  

\subsubsection{Opacity of Pluto's atmosphere at Lyman-$\alpha$ wavelengths}

On Pluto, solar ultraviolet light is dominated by Lyman-$\alpha$ photons, which control much of the photodissociation of CH$_4$ and of the subsequent hydrocarbon photochemistry \citep{Glad:16,Grun:18}. 
\citet{Bert:17} showed that the photochemical reactions are photon-limited in present-day Pluto's atmosphere, i.e. that enough gaseous CH$_4$ is present for all photons to be absorbed by CH$_4$ molecules.

Here we want to assess the opacity of Pluto's atmosphere at Lyman-$\alpha$ wavelengths over astronomical timescales. 
Indeed, if the atmospheric mixing ratio of CH$_4$ or the entire atmosphere collapsed in Pluto's past, then a direct photolysis of surface ices and tholins could have happened, which would help understanding the high degree of processing of the dark material in Cthulhu \citep{Grun:18}.

To do that, we first estimate the total incident flux of Lyman-$\alpha$ at Pluto over one orbit ($F_{tot}$), considering the solar as well as the interplanetary medium (IPM) Lyman-$\alpha$ sources \citep{Glad:15}, as given by Eq. 2 and Eq. 5 in \citet{Bert:17}:

\begin{equation}
F_{tot}(d_p) = \frac{F_{Earth}}{4 d_p^{2}}*0.875 + F_{IPM}
\label{Itot}
\end{equation}

We assume a constant solar Lyman-$\alpha$ flux at Earth $F_{Earth}$=$4\times10^{15}$ ph\,m$^{-2}$\,s$^{-1}$, a constant IPM flux at Pluto $F_{IPM}$=$7.25\times10^{11}$ ph\,m$^{-2}$\,s$^{-1}$ and a constant extinction factor of 0.875. The IPM flux does not strongly depend on the Sun-Pluto distance $d_p$ \citep{Glad:15}, therefore we consider it constant over time.
The integration of $F_{tot}$ over one Pluto orbit gives an annual mean incident Lyman-$\alpha$ flux of 1.3$\times10^{12}$ ph\,m$^{-2}$\,s$^{-1}$.
We then use Eq.~1 in B2018 to estimate the fraction of this incident Lyman-$\alpha$ flux reaching the surface (Beer’s law), by feeding this equation with the values of surface pressure and atmospheric CH$_4$ mixing ratio obtained from our simulations over 30 Myrs. 

\begin{figure}[!h]
\begin{center} 
	\includegraphics[width=15cm]{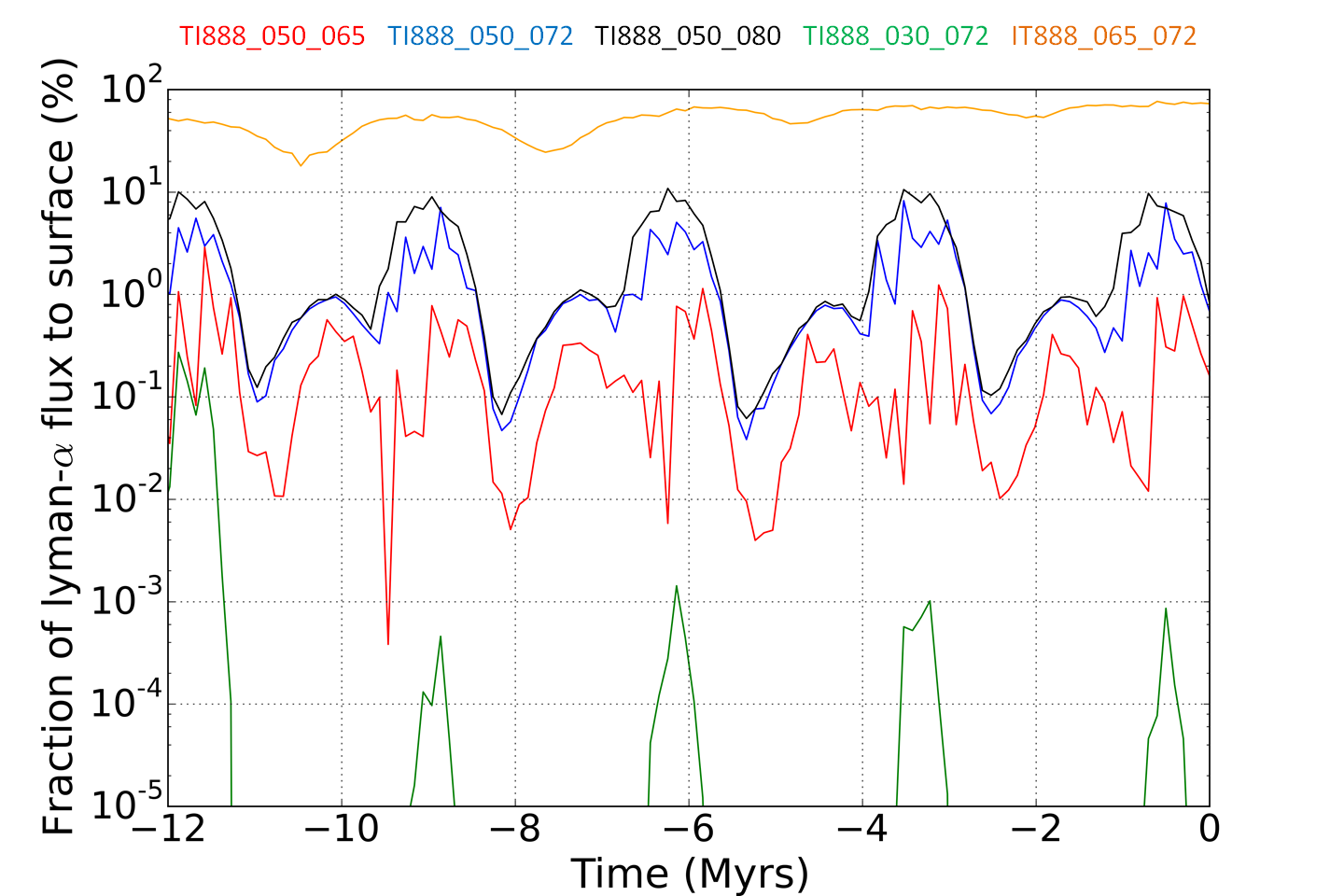}
\end{center} 
\caption{Fraction of Lyman-$\alpha$ flux reaching Pluto's surface (on annual average) for the same simulations than on \autoref{fig:vmr}.  \newline
} 
\label{fig:lym}
\end{figure}

\autoref{fig:lym} shows the results for the same simulations than those shown on \autoref{fig:vmr}. 
The atmosphere remains relatively opaque at Lyman-$\alpha$ wavelengths over astronomical timescales. For the most realistic simulations, the fraction of the annual mean incident flux that reaches the surface varies between 0.01 and 10$\%$ over time, with the lowest values obtained during high-obliquity periods, when mid-to-polar N$_2$ ice deposits are less stable. 
Over a current-year Pluto, we estimate that the fraction of Lyman-$\alpha$ reaching Pluto's surface is less than 1$\%$ of the total flux received.
These fractions of Lyman-$\alpha$ flux may be sufficient to have a significant effect on the chemistry of the N$_2$:CH$_4$:CO ice mixtures \citep{Mate:15,Grun:18}. Indeed, even if most of the Lyman-$\alpha$ flux is greatly attenuated most of the time, the photolysis of the ices goes on, albeit more slowly.

The fraction becomes negligible when the atmophere is enriched in gaseous CH$_4$, as it is the case if a low albedo of the equatorial CH$_4$ deposits is considered ($\#$TI888$\_$030$\_$072, green line). However, most of the incident flux can reach the surface if the atmospheric CH$_4$ mixing ratio is less than 0.01$\%$ over an entire year, as it is the case for the simulations where N$_2$ covers the equatorial CH$_4$ deposits ($\#$TI888$\_$065$\_$072, orange line). In this case, the CH$_4$ cycle is disrupted because the sources of gaseouse CH$_4$ are trapped by N$_2$ ice, and CH$_4$ can no longer block the energetic radiation, which would act directly on Pluto's surface ices.

\begin{sidewaystable}
\renewcommand{\arraystretch}{0.5} 
\centering
\caption{Settings and results of the simulations performed from 30~Myrs ago to present-day. From left to right, settings are: Run name (* indicates that the run is illustrated by figures in this paper), thermal inertia of N$_2$, CH$_4$, H$_2$O ice, equatorial and mid-to-polar CH$_4$ albedo. Results are: Loss of equatorial CH$_4$ ice after 30~Myrs, year of maximum pressure in current epoch, latitudes between which perennial N$_2$ ice deposits (pN$_2$) are obtained in the northern hemisphere, maximal thickness of these perennial deposits, same for seasonal deposits (sN$_2$), and maximal thickness of the equatorial N$_2$ ice deposits (formed on the modeled BTD).}
\label{tab:results}
\label{lasttable}
\begin{tiny} 
\tabcolsep=0.11cm
\begin{tabular}{m{3cm}m{1.3cm}m{1.3cm}m{1.3cm}m{1.3cm}m{1.8cm}m{1.3cm}m{1.3cm}m{1.3cm}m{1.3cm}m{1.3cm}m{1.3cm}m{1.3cm}}
\hline
\hline
\textbf{Name} & \textbf{TI$_{N_2}$} & \textbf{TI$_{CH_4}$} & \textbf{TI$_{H_2O}$} & A$_{CH_4}$ eq & A$_{CH_4}$ poles & L$_{CH_4}$ & Y$_p$ & Lat$_{pN_2}$ & Max$_{pN_2}$ & Lat$_{sN_2}$ & Max$_{sN_2}$ & Max$_{eqN_2}$ \\
 & \multicolumn{3}{c}{(J~s$^{-1/2}$~m$^{-2}$~K$^{-1}$)} &  & & (m) & & ($^\circ$) & (m) & ($^\circ$) & (m) & (m) \\
\hline
\hline
\textbf{$\#$TI888$\_$050$\_$060} & 800 & 800 & 800 & 0.5 & 0.6 & 75 & 1998.6 &  &  & 30-90 & 0.33 & 0 \\
\textbf{$\#$TI888$\_$050$\_$065*} & 800 & 800 & 800 & 0.5 & 0.65 & 87 & 1996.9 & 30  & 1.5 & 37.5-90 & 0.51 & 0 \\
\textbf{$\#$TI888$\_$050$\_$068} & 800 & 800 & 800 & 0.5 & 0.68 & 88 & 1995.1 & 30  & 4.0 & 37.5-90 & 0.59 & 0 \\
\textbf{$\#$TI888$\_$050$\_$072*} & 800 & 800 & 800 & 0.5 & 0.72 & 91 & 2000.4 & 30-37.5 & 4.9 & 45-90 & 0.75 & 0 \\
\textbf{$\#$TI888$\_$050$\_$080*} & 800 & 800 & 800 & 0.5 & 0.8 & 92 & 2017.9 & 30-37.5 & 7.3 & 45-90 & 0.90 & 0 \\
\textbf{$\#$TI888$\_$050$\_$065$\_$phase} & 800 & 800 & 800 & 0.5 & 0.65 & 72 & 2000.4 &   &  & 30-90 & 0.08 & 0 \\
\textbf{$\#$TI888$\_$050$\_$072$\_$phase*} & 800 & 800 & 800 & 0.5 & 0.72 & 86 & 1998.6 &   &  & 30-90 & 0.34 & 0 \\
\textbf{$\#$TI888$\_$050$\_$080$\_$phase} & 800 & 800 & 800 & 0.5 & 0.8 & 88 & 2017.9 &   &  & 30-90 & 0.77 & 0 \\
\textbf{$\#$TI888$\_$030$\_$065} & 800 & 800 & 800 & 0.3 & 0.65 & 1794 & 1996.9 & 30-37.5 & 3.2 & 45-90 & 0.54 & 0 \\
\textbf{$\#$TI888$\_$030$\_$072*} & 800 & 800 & 800 & 0.3 & 0.72 & 1776 & 2014.4 & 30-37.5 & 8.6 & 45-90 & 0.76 & 0 \\
\textbf{$\#$TI888$\_$060$\_$072*} & 800 & 800 & 800 & 0.6 & 0.72 & 10 & 1996.9 & 30 & 4.9 & 37.5-90 & 0.66 & 77 \\
\textbf{$\#$TI888$\_$060$\_$080} & 800 & 800 & 800 & 0.6 & 0.8 & 6 & 2000.4 & 30-37.5 & 7.7 & 45-90 & 0.86 & 97 \\
\textbf{$\#$TI888$\_$065$\_$072} & 800 & 800 & 800 & 0.65 & 0.72 & 3 & 2002.1 &   &  & 52.5-90 & 0.11 & 357 \\
\textbf{$\#$TI888$\_$065$\_$080} & 800 & 800 & 800 & 0.65 & 0.8 & 1 & 1996.9 & 30-37.5 & 4.7 & 45-90 & 0.68 & 185 \\
\hline                       
\textbf{$\#$TI884$\_$050$\_$065*} & 800 & 800 & 400 & 0.5 & 0.65 & 85 & 2014.4 & 30 & 10.5 & 22.5-90 & 0.51 & 0 \\
\textbf{$\#$TI884$\_$050$\_$072*} & 800 & 800 & 400 & 0.5 & 0.72 & 90 & 2019.6 & 30-37.5 & 6.2 & 22.5-90 & 0.81 & 0 \\
\textbf{$\#$TI8812$\_$050$\_$065*} & 800 & 800 & 1200 & 0.5 & 0.65 & 85 & 1995.1 & 30 & 3.8 & 37.5-90 & 0.44 & 0 \\
\textbf{$\#$TI8812$\_$050$\_$072*} & 800 & 800 & 1200 & 0.5 & 0.72 & 91 & 2014.4 & 30-37.5 & 14.6 & 45-90 & 0.70 & 0 \\
\textbf{$\#$TI848$\_$050$\_$065} & 800 & 400 & 800 & 0.5 & 0.65 & 106 & 1996.9 & 30 & 5.5 & 37.5-90 & 0.58 & 78 \\
\textbf{$\#$TI848$\_$050$\_$072} & 800 & 400 & 800 & 0.5 & 0.72 & 106 & 2005.6 & 30-37.5 & 9.4 & 45-90 & 0.78 & 72 \\
\textbf{$\#$TI8128$\_$050$\_$065} & 800 & 1200 & 800 & 0.5 & 0.65 & 67 & 1996.9 & 30 & 2.0 & 37.5-90 & 0.47 & 0 \\
\textbf{$\#$TI8128$\_$050$\_$072} & 800 & 1200 & 800 & 0.5 & 0.72 & 73 & 1998.6 & 30 & 11.2 & 37.5-90 & 0.71 & 0 \\
\textbf{$\#$TI488$\_$050$\_$065} & 400 & 800 & 800 & 0.5 & 0.65 & 86 & 1995.1 & 30 & 2.4 & 37.5-90 & 0.51 & 0 \\
\textbf{$\#$TI488$\_$050$\_$072} & 400 & 800 & 800 & 0.5 & 0.72 & 90 & 2002.1 & 30-37.5 & 5.5 & 45-90 & 0.77 & 0 \\
\textbf{$\#$TI1288$\_$050$\_$065} & 1200 & 800 & 800 & 0.5 & 0.65 & 87 & 1995.1 & 30 & 2.8 & 37.5-90 & 0.57 & 0 \\
\textbf{$\#$TI1288$\_$050$\_$072} & 1200 & 800 & 800 & 0.5 & 0.72 & 92 & 2000.4 & 30-37.5 & 6.2 & 45-90 & 0.74 & 0 \\
\textbf{$\#$TI444$\_$050$\_$072} & 400 & 400 & 400 & 0.5 & 0.72 & 105 & 2010.9 & 30-37.5 & 8.3 & 22.5-90 & 0.77 & 91 \\
\textbf{$\#$TI444$\_$050$\_$080} & 400 & 400 & 400 & 0.5 & 0.8 & 100 & 2016.1 & 30-37.5 & 6.7 & 22.5-90 & 0.93 & 105 \\
\textbf{$\#$TI844$\_$050$\_$060} & 800 & 400 & 400 & 0.5 & 0.6 & 105 & 2010.9 & 30 & 7.3 & 22.5-90 & 0.46 & 22 \\
\textbf{$\#$TI844$\_$050$\_$065} & 800 & 400 & 400 & 0.5 & 0.65 & 108 & 2014.4 & 30 & 10.9 & 22.5-90 & 0.56 & 82 \\
\textbf{$\#$TI844$\_$050$\_$072} & 800 & 400 & 400 & 0.5 & 0.72 & 108 & 2014.4 & 30-37.5 & 10.2 & 22.5-90 & 0.79 & 83 \\
\textbf{$\#$TI8412$\_$050$\_$065} & 800 & 400 & 1200 & 0.5 & 0.65 & 133 & 1996.9 & 30 & 6.3 & 37.5-90 & 0.56 & 72 \\
\textbf{$\#$TI8412$\_$050$\_$072} & 800 & 400 & 1200 & 0.5 & 0.72 & 108 & 1995.1 & 30-37.5 & 15.5 & 45-90 & 0.73 & 82 \\
\textbf{$\#$TI8124$\_$050$\_$060} & 800 & 1200 & 400 & 0.5 & 0.6 & 59 & 2007.4 &   &  & 30-90 & 0.31 & 0 \\
\textbf{$\#$TI8124$\_$050$\_$065} & 800 & 1200 & 400 & 0.5 & 0.65 & 66 & 2012.6 & 30 & 5.0 & 22.5-90 & 0.49 & 0 \\
\textbf{$\#$TI8124$\_$050$\_$072} & 800 & 1200 & 400 & 0.5 & 0.72 & 72 & 2019.6 & 30-37.5 & 6.4 & 22.5-90 & 0.81 & 0 \\
\textbf{$\#$TI81212$\_$050$\_$065} & 800 & 1200 & 1200 & 0.5 & 0.65 & 66 & 1995.1 & 30 & 2.1 & 37.5-90 & 0.50 & 0 \\
\textbf{$\#$TI81212$\_$050$\_$072} & 800 & 1200 & 1200 & 0.5 & 0.72 & 71 & 1996.9 & 30 & 6.1 & 37.5-90 & 0.63 & 0 \\
\hline
\end{tabular}
\end{tiny}
\end{sidewaystable}

\section{Discussions} 
\label{sec:discuss}

In this section, we first compare our results with Pluto's observations and further discuss the possible scenarios for the formation and evolution of the N$_2$ and CH$_4$ reservoirs. 

\subsection{Comparison of our results with Pluto's observations}
\label{sec:comp}

\subsubsection{The massive equatorial CH$_4$ deposits}

In our model, CH$_4$ ice spontaneously tends to accumulate in the equatorial regions over astronomical timescales (see Section~\ref{sec:uniform}), forming thick perennial CH$_4$ deposits. This explains the presence of the massive equatorial CH$_4$-rich Bladed Terrain Deposits observed by New Horizons \citep{Moor:16,Moor:18,Earl:18b}. 

We obtain a slightly larger accumulation of CH$_4$ ice north of the equator, which is consistent with the observed extension of the BTD, ranging in latitude from about 5$^{\circ}$S to 28$^{\circ}$N \citep[see Fig. 4 in][]{Earl:18b,Moor:18}. This location is reproduced by our simulations when assuming a medium-to-high soil thermal inertia (\autoref{fig:refuniform}).  
Similar results are obtained by starting the same simulation at another epoch. The equatorial deposits are also slightly more extended to the north if we start the simulation at $t_0=$-100~Myrs, but to the south if we start at $t_0=$-200 or -300~Myrs, which are assumed to be different epochs with a perihelion occurring at northern fall or winter during maximum obliquity periods (Figure 5 in B2018). 

Our results show that 30~Myrs is too short a period of time to form km-thick equatorial CH$_4$-rich deposits like the BTD.
The BTD may be relatively old, as suggested by their dark albedo ($\sim$~0.5, \citet{Bura:17}). The lack of craters suggests an upper limit on their age of 300~Myrs \citep{Moor:18}, although ancient craters may have been erased by intense sublimation of these terrains.   
We estimate that $\sim$~1~km thick CH$_4$-rich deposits could form in the equatorial regions (30$^\circ$S-30$^\circ$N) over 50-100~Myrs, assuming an initial reservoir elsewhere (e.g. at the poles as in Section~\ref{sec:uniform}). Once this reservoir is depleted, extra hundreds of Myrs would be necessary to collect CH$_4$ ice from the edges of the equatorial regions toward more equatorial latitudes and to pile up km-thick amounts of ice. If the BTD are only few hundreds meter thick, they could have formed over 50~Myrs or less.

In the model, the equatorial CH$_4$ deposits form at all longitudes outside Sputnik Planitia. In reality, on Pluto, the unusual color of these terrains is only seen between longitudes 210$^{\circ}$E to 40$^{\circ}$E \citep{Olki:17,Moor:18,Earl:18b}, and ground-based observations support the presence of more CH$_4$-rich deposits at these longitudes \citep{GrunBuie:01,Grun:13}. This longitudinal asymmetry must be investigated with a full 3D Global Climate Model since it may be due to a dynamical effect of Pluto's atmosphere. 

We also show in Section~\ref{sec:bladed} that if the mid-to-polar CH$_4$ deposits are bright enough (we evaluate the critical CH$_4$ albedo around A$_{CH4}$=0.6), then mm-thick seasonal or meter-thick perennial N$_2$ deposits can form in these regions, preferentially at low elevations.
In this case, we find that the BTD are not stable and correspond to a net sublimation zone at the astronomical timescale, because CH$_4$ ice is transferred from the BTD toward the mid-latitudes regions where it remains trapped by the perennial or seasonal N$_2$ deposits.
Assuming a CH$_4$ albedo of 0.5 for the modeled BTD, we obtain a loss L$_{CH_4}\sim$100~m of ice over the 30~Myrs in most of our simulations (see Table~\ref{tab:results}). The km-thick BTD could therefore disappear within 300~Myrs.

These results suggest that the BTD were thicker in the past, and are now gradually disappearing. 
This is to be related to their “bladed” morphology, thought to be controlled by sublimation process \citep{Moores:17b,Moor:17,Moor:18}.

\subsubsection{The perennial reservoirs of N$_2$ ice}

Our simulations show that, the closer to the equator, the more perennial are the N$_2$ reservoirs (outside of Sputnik Planitia).
As a general trend, our model simulates two types of perennial N$_2$ reservoirs (apart from Sputnik Planitia). First, up to 200-300~m thick N$_2$ ice deposits can form at the equator, at low elevations where relatively bright CH$_4$ ice remains (albedo $\sim$0.6). This result is consistent with the detection of N$_2$ ice in East Tombaugh Regio, in the depressions surrounding the BTD and in some deep craters in Cthulhu (e.g. Elliot crater, where evidences of polygonal cells may be supportive of a thick deposit), and suggests that these deposits may be stable over several Myrs. 
Second, a 10-20~m-thick mid-latitudinal band (30-45$^\circ$N) of N$_2$ ice forms in most of our simulations and remains over several Myrs, which is consistent with observations \citep[see Fig. 15 in][]{Schm:17, Prot:17}. The extent of such N$_2$ ice deposits toward lower latitudes is limited by the  presence of tholin-covered regions, which tend to darken the surrounding areas thus preventing further accumulation of volatile ice.  

All N$_2$ ice deposits are balanced by the main reservoir in Sputnik Planitia: as they form, the reservoir and therefore the surface level of N$_2$ ice within the basin decreases which reinforces the N$_2$ condensation in the basin \citep{BertForg:16}, thus limiting the amount of N$_2$ ice that can form elsewhere. We also note that CH$_4$-rich ice on Pluto may not be trapped and buried by very large N$_2$-rich deposits, since N$_2$ ice is twice as dense as that of CH$_4$ and may sink through the CH$_4$ ice. For instance, on Triton, the Cantaloupe terrain may result from such a process \citep{ScheJack:93}.

\subsubsection{The mid-to-polar N$_2$ and CH$_4$ deposits}


At the north pole, New Horizons detected CH$_4$-rich deposits \citep{Grun:16,Schm:17,Prot:17}. Most of our simulations reproduce this trend by predicting a seasonal (mm-thick) CH$_4$ ice deposit exposed on Pluto's surface at the north pole in 2015 (e.g. \autoref{fig:refbladedseason},\autoref{fig:poles065TIH2O400}), quickly disapearing and revealing the dark volatile-free surface during the following years.
However, such a thin deposit should have allowed the spectrometers on-board New Horizons for the detection of the water ice bedrock below. The fact that water ice has not been detected anywhere in the polar region strongly suggests that at least several centimeters or even meters of CH$_4$-rich ice cover this region (or else the water ice is buried under photolytic byproducts).

Simulations $\#$TI888$\_$030$\_$072 and $\#$TI888$\_$050$\_$080 (\autoref{fig:bladed03} and \autoref{fig:poles065} respectively) show that such perennial CH$_4$-rich deposits can be obtained at the poles, by assuming a very dark albedo for the BTD or a very bright albedo for the polar CH$_4$ deposits. 
These results may be related to the yellow aspect of the polar region in false-color images \citep{Olki:17}. For instance, the color may be due to the presence of tholins mixed with (or seen through) the thin CH$_4$-rich frost \citep{Grun:18}. Alternatively, the polar reservoir of CH$_4$ may be perennial and have accumulated tholins over the last Myrs. The higher concentration of tholins in these deposits may be responsible for their peculiar color.

At northern mid-latitudes, we obtain the formation of N$_2$ ice deposits (on the bright CH$_4$ deposits), which tend to be seasonal above 37$^\circ$N as they sublime from the poles during end spring/summer. 
This result explains the mid-latitudinal band of N$_2$-rich ice observed in 2015 \citep{Prot:17,Schm:17}, and is consistent with the latitudinal distribution of the different ice mixtures observed at these latitudes, indicative of N$_2$:CO sublimation processes \citep{Prot:17,Schm:17}.
Our results predict that N$_2$ ice sublimation will continue during northern spring/summer (with a sublimation front advancing southward, as suggested in \citet{Prot:17}) and reveal more CH$_4$-rich terrains, while the redeposition of N$_2$ ice will occur in the southern winter hemisphere.
Note that N$_2$ ice is always more stable in the depressions than outside, which explains the patchy distribution of N$_2$ found at northern mid-latitudes \citep[see Fig. 15 in ][]{Schm:17}. 

In the model, the seasonal mid-latitudinal N$_2$ ice deposits are able to trap large amounts of CH$_4$ ice, provided that a constant source is present elsewhere (in this case, the BTD). This also occurs in simulations performed at other epochs ($t_0$=-100, -200, -300~Myrs) and explains the thick mantle of ice observed by New Horizons at these latitudes.

In the model, the N$_2$ ice deposits forming between 30$^\circ$N-37$^\circ$N tend to be perennial, because they are located in colder regions on average over several Myrs (this is an effect of subsurface thermal inertia, as shown by Fig. 4.b in B2018). 
At lower latitudes, the dark albedo of the equatorial volatile-free surface prevent further N$_2$ condensation. 
However, thin CH$_4$ frosts may form on these dark terrains during winter (see e.g \autoref{fig:poles065TIH2O400}). This is consistent with the third latitudinal band observed on Pluto and mentioned in \citet{Prot:17}, where CH$_4$-rich deposits are observed between 20$^\circ$N-30$^\circ$N \citep[also shown in][]{Earl:18b}. At these latitudes, which border the dark tholin-covered regions, N$_2$ ice deposits are not stable due to the darker albedo and the high rate of contamination and darkening by tholins. 
However, CH$_4$-rich frosts may form and last until spring/summer or even last longer and form perennial deposits.

Note that the exact latitudes where the volatile ice deposits form in the model are sensitive to the surface properties (albedo, emissivity, thermal inertia). In addition, dilution and mixing processes of these ices should impact their latitudinal distribution. 

\subsubsection{Best case simulations}

A simulation that best matches Pluto's observation would show (1) an ice distribution in 2015 as observed by New Horizons, that is with CH$_4$ ice exposed at the north pole and N$_2$ ice subliming at mid-latitudes, (2) a peak surface pressure occurring after 2015, and (3) an atmospheric mixing ratio for CH$_4$ around 0.5$\%$ in the period 2010-2015 \citep{Lell:11a,Lell:15}. 
In this paper, although we explored many cases and obtained a plethora of results, it is difficult to find one simulation that reconciles all these observations. Generally speaking, simulations performed with a thermal inertia for water ice of 400~SI give the best results. This thermal inertia is low compared to the estimations \citep{Leyr:16} but allows for thicker deposits of CH$_4$ at the poles, which is consistent with the observations.   

\subsection{Scenario for the formation of observed CH$_4$-rich and N$_2$-rich reservoirs}

\begin{figure}[!h]
\begin{center} 
	\includegraphics[width=15cm]{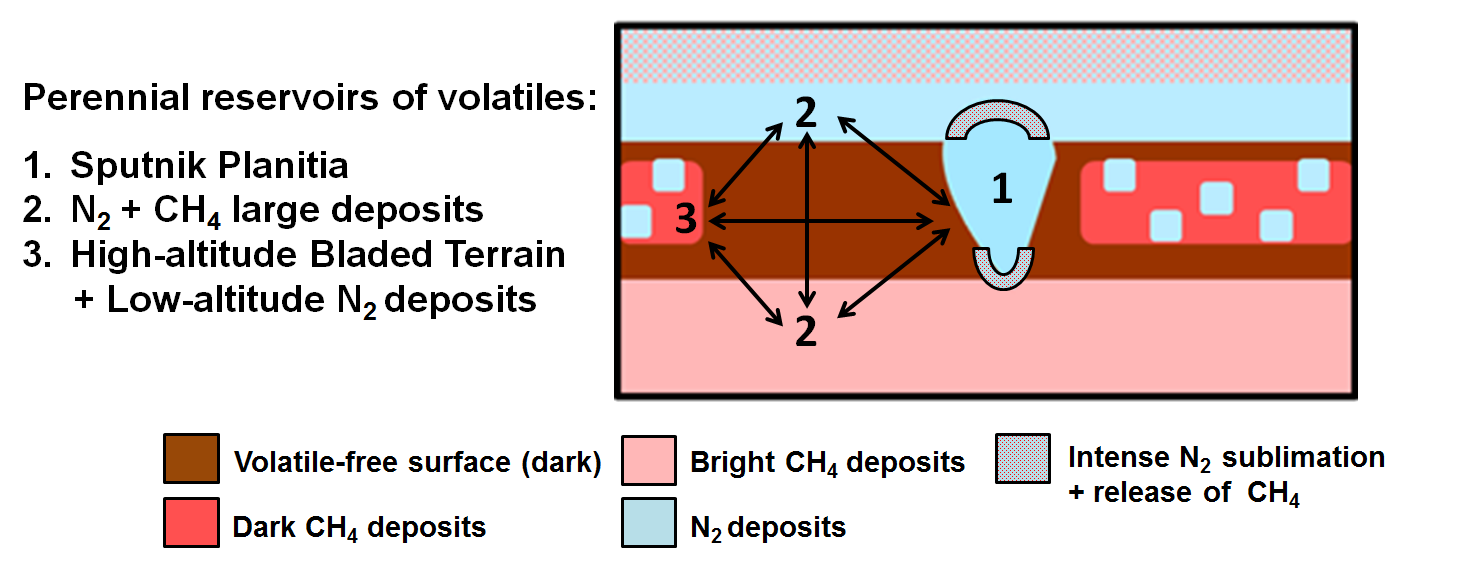}
\end{center} 
\caption{The perennial reservoirs of N$_2$ and CH$_4$ ice identified on Pluto. The black arrows represent the possible exchanges of volatile ice that may occur over timescales of millions of years.  
} 
\label{fig:reservoirs}
\label{lastfig}
\end{figure}

We have identified two perennial reservoirs of volatile ice on Pluto which add to Sputnik Planitia, as shown in \autoref{fig:reservoirs}: the mid-latitudinal regions (mantle of CH$_4$ ice and meter-thick N$_2$ ice deposits) and the equatorial regions east of Sputnik Planitia (BTD of CH$_4$ ice and low-altitude N$_2$ ice deposits). 

Our model is able to reproduce the formation of the equatorial and mid-latitudinal perennial CH$_4$ reservoirs, although not at the same time. Indeed, both the BTD and the mid-latitudinal CH$_4$ mantle need a source of available CH$_4$ ice to form, and in our simulations one reservoir dominates the other, depending on the assumed CH$_4$ albedo.
This suggests a complex history for the formation of these perennial reservoirs.

The reservoir in Sputnik Planitia must have been the first to form, since the basin is the oldest geologic feature identified on Pluto \citep[\textgreater~4~billion years]{Moor:16}, and since its infilling with all available surface N$_2$ ice has been modeled and estimated to be complete by tens of millions of years following its formation \citep{Bert:18}. 

We could imagine that at a time in Pluto's history, large CH$_4$-rich ice deposits formed at the equator.
At first, they may have been covered by N$_2$-rich ice deposits, but as they thickened up and reached higher altitudes with time, N$_2$-rich ice probably became less stable on these deposits and remained in depressions only. The high-altitude CH$_4$ ice deposits may have then become older and darker with time, forming a plateau of darkened CH$_4$-rich, precursor to the BTD. 
In our simulations, we show that there is a net transport of CH$_4$ ice from the BTD to the mid-latitudes, suggesting that the thick mantle of CH$_4$ ice is subsequent to the BTD, which could have formed by erosionnal sublimation.
Consequently, if the BTD formed very early in Pluto's history, why have they not entirely disappeared by now? 
One solution is that the current BTD formed only recently in Pluto's history, and may be currently disappearing. One could imagine that the astronomical cycles of Pluto may have changed over the last billion years and created the conditions for the BTD only recently. However, Pluto is thought to be subject to very little perturbations \citep{Dobr:97} and the presence of non-eroded ancient craters at the equator demonstrates a certain stability of the astronomical cycles \citep{Binz:17}. Another solution is that a long-term cycle exists between the perennial reservoirs, as illustrated by the black arrows on \autoref{fig:reservoirs}. 
In this case, processes not taken into account in our model must exist and refill Pluto's system with CH$_4$ gas and ice. 

For instance, large amounts of CH$_4$ could be released at the northern and southern edges of the Sputnik Planitia ice sheet during the high obliquity periods, where intense N$_2$ sublimation occurs \citep{Bert:18}. In fact, New Horizons observations of the northern edge of the ice sheet revealed very dark plains of N$_2$-rich ice enriched in CH$_4$ ice (1-2$\%$, compared to 0.3$\%$ in the rest of the ice sheet), which supports this scenario \citep{Prot:17}. 

Albedo and topography run-away variations may also play a significant role in redistributing N$_2$ and CH$_4$ ice to different reservoirs \citep{Earl:18}, as well as changes in ice composition, saturation, contamination or irradiation \citep{Schm:17,Prot:17,Grun:18}. Another possibility is the re-supply of large amounts of CH$_4$ from Pluto's interior, where sources of CH$_4$ clathrate could be stored and released over time via outgassing or cryovolcanism, as proposed on Titan \citep{LuniAtre:08,Moor:16}.

The lack of knowledge about these mechanisms makes it difficult to infer the total CH$_4$ inventory in the system. A lower limit could be estimated by assuming that the high latitudes are only covered by thin layers of CH$_4$ ice and that most of the CH$_4$ reservoir is contained in the Bladed Terrain Deposits (the contribution of 0.3-0.5$\%$ of CH$_4$ in the N$_2$ reservoir of Sputnik Planitia is negligible). Such a reservoir has been estimated to 22~m on global average in Section \ref{sec:res}, for 500~m thick BTD, but could be raised to 100~m if we assume 2-2.5~km thick BTD. An upper limit of 1~km could be estimated by considering the BTD and a 1-km thick CH$_4$ mantle at latitudes higher than 15$^\circ$ (740~m on global average).  
Note that the weak escape rate of CH$_4$ observed by New Horizons suggests that the reservoir of CH$_4$ changed by only 28~m over the age of the solar system \citep{Glad:16}.  

\section{Conclusions}
\label{sec:conclusion}

The Pluto volatile transport model has been used to simulate the evolution of large N$_2$ and CH$_4$ ice reservoirs over seasonal and astronomical timescales, in response to the Milankovitch paleoclimate cycles.
This complements the work done by B2018, which only explored the cycles of N$_2$. 

Our simulations reproduce the formation of massive perennial deposits of CH$_4$-rich ice at the equator, explaining the observation of the Bladed Terrain Deposits at these locations. The configuration of Pluto's orbit and obliquity is responsible for the small but detectable asymmetry of these terrains around the equator, which form further in the north than in the south, as seen in our model. 

We demonstrate that high CH$_4$ ice albedo values may sufficiently cool the surface and trigger N$_2$ condensation which then cold traps more CH$_4$ ice. 
We obtained a plethora of results depending on the assumed albedo for CH$_4$ ice, which controls the perennial or seasonal nature of the deposits.
Assuming relatively dark Bladed Terrain Deposits allows the formation of perennial N$_2$ ice in the depressions of these terrains only, which is consistent with the observations. 
Assuming relatively bright mid-to-polar CH$_4$ ice deposits leads to the formation of up to 10-m thick N$_2$ deposits on the CH$_4$ ice deposits, in particular at low elevations. During northern spring and summer, these N$_2$ deposits sublime from the pole, explaining the latitudinal distribution of N$_2$-rich and CH$_4$-rich ices observed by New Horizons in 2015 at the mid-northern latitudes \citep{Prot:17,Schm:17}.

At the north pole, the disapearance of CH$_4$-rich frost during the next decade predicted by most of our simulation may not happen because the observations suggest larger deposits than those simulated. However, this is a testable prediction, as well as the removal of N$_2$-rich ice from the northern high latitudes, since it should result in changes in Pluto's spectrum and maybe albedo and color, observable from Earth.

Our simulations also show that a large amount of CH$_4$ ice can accumulate in the mid-latitudes due to the cold-trap effect of N$_2$ ice, forming a thick mantle consistent with the observations of Pluto's surface by New Horizons. 
In the model, CH$_4$ ice is moved from the Bladed Terrain Deposits, which tends to disappear as the mid-latitude mantle forms.

These simulations indicate that there is always enough gaseous CH$_4$ in Pluto's atmosphere for it to remain relatively opaque at Lyman-$\alpha$ wavelengths over astronomical cycles, in particular during high-to-moderate obliquity periods. However, the small amounts of Lyman-$\alpha$ flux reaching the surface can be significant to the surface chemistry.

Finally, our results highlight the strong coupling between the CH$_4$ and the N$_2$ cycle and the role of CH$_4$ material as a controlling agent of this coupling. They suggest a complex history for Pluto's perennial reservoirs as large amounts of CH$_4$ ice may have been exchanged between Sputnik Planitia, the Bladed Terrain, and the mid-latitudes, over timescales of hundreds of million years.
The evolution of these reservoirs may be driven by positive (``run-away``) or negative feedbacks involving ice albedo, emissivity, dilution and mixing coefficient, contamination by haze particles, or ice irradiation.
In order to improve our understanding of Pluto's surface, future versions of the Pluto volatile transport model should implement these processes and explore how they impact Pluto's climate.

\newpage
\bibliographystyle{plainnat}
\bibliography{biblio}

\ack
This work was supported by the CNES. It is based on observations of the New Horizons space mission.
The authors thank the whole NASA \textit{New Horizons} instrument and scientific team for their excellent work on a fantastic mission and their interest in this research. 
T. B. was supported for this research by an appointment to the National Aeronautics and Space Administration (NASA) Post-doctoral Program at the Ames Research Center administered by Universities Space Research Association (USRA) through a contract with NASA.
\label{lastpage}


\clearpage 

\end{document}